\documentclass[a4paper,12pt]{article}

\usepackage{epsfig}
\usepackage{amssymb}
\usepackage{graphicx}
\usepackage{latexsym}
\usepackage{bbold}
\usepackage{bbm}
\usepackage[mathscr]{eucal}
\usepackage[english]{babel}

\def\half{\frac{1}{2}} 
\def\ra{\rightarrow}

\usepackage{color}
\definecolor{rosso}{cmyk}{0,1,1,0.4}
\definecolor{rossos}{cmyk}{0,1,1,0.55}
\definecolor{rossoc}{cmyk}{0,1,1,0.2}
\definecolor{blu}{cmyk}{1,1,0,0.3}
\definecolor{blus}{cmyk}{1,1,0,0.6}
\definecolor{bluc}{cmyk}{1,1,0,0.1}
\definecolor{verde}{cmyk}{0.92,0,0.59,0.25}
\definecolor{verdec}{cmyk}{0.92,0,0.59,0.15}
\definecolor{verdes}{cmyk}{0.92,0,0.59,0.4}

\def\circa#1{\,\raise.3ex\hbox{$#1$\kern-.75em\lower1ex\hbox{$\sim$}}\,}

\newcommand{\eV}{\,{\rm eV}}

\makeatletter
\newcommand{\beq}{\begin{equation}}
\newcommand{\eeq}{\end{equation}}
\newcommand{\bea}{\begin{eqnarray}}
\newcommand{\eea}{\end{eqnarray}}
\newcommand{\ba}{\begin{array}}
\newcommand{\ea}{\end{array}}
\newcommand{\bn}{\begin{enumerate}}
\newcommand{\en}{\end{enumerate}}
\newcommand{\bc}{\begin{center}}
\newcommand{\ec}{\end{center}}

\newcommand{\gsim}{\lower.7ex\hbox{$\;\stackrel{\textstyle>}{\sim}\;$}}
\newcommand{\lsim}{\lower.7ex\hbox{$\;\stackrel{\textstyle<}{\sim}\;$}}
\newcommand{\baz}{\begin{array}{cc}}
\newcommand{\bad}{\begin{array}{ccc}}
\def\gtap{\mathrel{ \rlap{\raise 0.511ex \hbox{$>$}}{\lower 0.511ex
   \hbox{$\sim$}}}} 
\def\ltap{\mathrel{ \rlap{\raise 0.511ex
   \hbox{$<$}}{\lower 0.511ex \hbox{$\sim$}}}}
   \newcommand{\deltaatm}{\mbox{$\Delta m^2_{\mathrm{A}}$}}
   \newcommand{\deltasol}{\mbox{$ \Delta m^2_{\odot}$}}
\newcommand{\dmsol}{\mbox{$\Delta m^2_{\odot}$}}
\newcommand{\dma}{\mbox{$\Delta m^2_{\rm A}$}}

\newcommand{\pmns}{\mbox{$ U$}}

\renewcommand{\thefootnote}{\fnsymbol{footnote}}

\begin{document}
\begin{titlepage}
\hfill
\vbox{
    \halign{#\hfil        \cr
      SISSA 60/2007/EP \cr
      arXiv:0709.0413\cr}}  
\vspace*{10mm}
  \begin{center}

{\large \bf Effects of Lightest Neutrino Mass in 
Leptogenesis \\[0.3cm]
}
%
\vspace*{7mm}

{\bf E.~Molinaro\footnote[1]{E-mail: molinaro@sissa.it}, S.~T.~Petcov\footnote[2]{Also at: Institute
of Nuclear Research and Nuclear Energy, Bulgarian Academy 
of Sciences, 1784 Sofia, Bulgaria.},
T.~Shindou\footnote[3]{E-mail: tetsuo.shindou@desy.de}$^{,}$\footnote{Current 
address: Deutsches Elektronen-Synchrotron DESY, Notkestra{\ss}e 85, 
D-22603 Hamburg, Germany.} and Y.~Takanishi\footnote[4]{E-mail: yasutaka@sissa.it}}
 
\vskip 6pt

{\it SISSA and INFN-Sezione di Trieste, Trieste I-34014, Italy}\\

\vspace{0.6cm}
\end{center}
\begin{abstract}
\noindent The effects of the lightest neutrino mass in ``flavoured''
leptogenesis are investigated in the case when the CP-violation
necessary for the generation of the baryon asymmetry of the Universe
is due exclusively to the Dirac and/or Majorana phases in the neutrino
mixing matrix $U$.  The type I see-saw scenario with three heavy
right-handed Majorana neutrinos having hierarchical spectrum is
considered.  The ``orthogonal'' parametrisation of the matrix of
neutrino Yukawa couplings, which involves a complex orthogonal matrix
$R$, is employed. Results for light neutrino mass spectrum with normal
and inverted ordering (hierarchy) are obtained.  It is shown, in
particular, that if the matrix $R$ is real and CP-conserving and the
lightest neutrino mass $m_3$ in the case of inverted hierarchical
spectrum lies the interval $5\times 10^{-4}~{\rm eV} \ltap m_3 \ltap
7\times 10^{-3}~{\rm eV}$, the predicted baryon asymmetry can be
larger by a factor of $\sim 100$ than the asymmetry corresponding to
negligible $m_3 \cong 0$. As consequence, we can have successful
thermal leptogenesis for $5\times 10^{-6}~{\rm eV} \ltap m_3 \ltap
5\times 10^{-2}$ eV even if $R$ is real and the only source of
CP-violation in leptogenesis is the Majorana and/or Dirac phase(s) in
$U$.
\end{abstract}

\begin{itemize}
\item[] PACS numbers: 98.80.Cq, 14.60.Pq, 14.60.St
\item[] keywords: thermal leptogenesis, seesaw mechanism, lepton flavour effects
\end{itemize}

\end{titlepage}

\renewcommand{\thefootnote}{\arabic{footnote}}
\setcounter{footnote}{0}
\setcounter{page}{1}

\newpage

\section{Introduction}
\indent\ In the present article we continue to investigate the
possible connection between leptogenesis~\cite{FY,kuzmin} (see also,
e.g.~\cite{LG1,LG2}) and the low energy CP-violation in the lepton
(neutrino) sector (for earlier discussions see,
e.g.~\cite{others,others2,PPR03,PRST05} and the references quoted
therein).  It was shown recently~\cite{PPRio106} that the CP-violation
necessary for the generation of the observed baryon asymmetry of the
Universe in the thermal leptogenesis scenario can be due exclusively
to the Dirac and/or Majorana CP-violating phases in the
Pontecorvo-Maki-Nakagawa-Sakata (PMNS)~\cite{BPont57} neutrino mixing
matrix, and thus can be directly related to the low energy
CP-violation in the lepton sector (e.g. in neutrino oscillations,
etc.).  The analysis performed in~\cite{PPRio106} (see
also~\cite{BrancoJ06,SBDibari06}) was stimulated by the progress made
in the understanding of the importance of lepton flavour effects in
leptogenesis~\cite{Barbieri99,Nielsen02,davidsonetal,niretal,davidsonetal2,antusch}.
It led to the realisation that these effects can play crucial role in
the leptogenesis scenario of baryon asymmetry
generation~\cite{davidsonetal,niretal,davidsonetal2}.  It was noticed
in~\cite{niretal}, in particular, that ``Scenarios in which
$\epsilon_1 = 0$ while $\epsilon_1^j \neq 0$ entail the possibility
that the phases in the light neutrino mixing matrix $U$ are the only
source of CP violation.'', $\epsilon_1^j$ and $\epsilon_1$ being
respectively the individual lepton number and the total lepton number
CP violating asymmetries.

  As is well-known, the leptogenesis theory~\cite{FY} is based on the
see-saw mechanism of neutrino mass generation~\cite{seesaw}.  The
latter provides a natural explanation of the observed smallness of
neutrino masses (see, e.g.~\cite{STPNu04,MoscowH3Mainz,Hann06}).  An
additional appealing feature of the see-saw scenario is that through
the leptogenesis theory it allows to relate the generation and the
smallness of neutrino masses with the generation of the baryon
(matter-antimatter) asymmetry of the Universe, $Y_B$.

The non-supersymmetric version of the type I see-saw model with two or
three heavy right-handed (RH) Majorana neutrinos is the minimal scheme
in which leptogenesis can be implemented.  In~\cite{PPRio106} the
analysis was performed within the simplest type I see-saw mechanism of
neutrino mass generation with three heavy RH Majorana neutrinos,
$N_j$, $j=1,2,3$.  Taking into account the lepton flavour effects in
leptogenesis it was shown \cite{PPRio106}, in particular, that if the
heavy Majorana neutrinos have a hierarchical spectrum, i.e. if $M_1
\ll M_{2,3}$, $M_j$ being the mass of $N_j$, the observed baryon
asymmetry $Y_B$ can be produced even if the only source of
CP-violation is the Majorana and/or Dirac phase(s) in the PMNS
matrix\footnote{The same result was shown to hold also for
  quasi-degenerate in mass heavy RH Majorana neutrinos
  \cite{PPRio106}.}  
$U_{\rm PMNS}\equiv U$. Let us recall that in the
case of hierarchical spectrum of the heavy Majorana neutrinos, the
lepton flavour effects can be significant in leptogenesis provided the
mass of the lightest one $M_1$ satisfies the constraint
\cite{davidsonetal,niretal,davidsonetal2} (see also \cite{DiBGRaff06}): $M_1
\ltap 10^{12}~{\rm GeV}$.  In this case the predicted value of the
baryon asymmetry depends explicitly (i.e.  directly) on $U$ and on the
CP-violating phases in $U$.  The results quoted above were
demonstrated to hold both for normal hierarchical (NH) and inverted
hierarchical (IH) spectrum of masses of the light Majorana neutrinos
(see, e.g. \cite{STPNu04}).  In both these cases they were obtained
for negligible lightest neutrino mass and CP-conserving elements of
the orthogonal matrix $R$, present in the ``orthogonal''
parametrisation \cite{Casas01} of the matrix of neutrino Yukawa
couplings.  The CP-invariance constraints imply \cite{PPRio106} that
the matrix $R$ could conserve the CP-symmetry if its elements are real
or purely imaginary.  As was demonstrated in \cite{PPRio106}, for NH
spectrum and negligible lightest neutrino mass $m_1$ one can have
successful thermal leptogenesis with real $R$.  In contrast, in the
case of IH spectrum and negligible lightest neutrino mass ($m_3$), the
requisite baryon asymmetry was found to be produced for CP-conserving
matrix $R$ only if certain elements of $R$ are purely imaginary: for
real $R$ the baryon asymmetry $Y_B$ is strongly suppressed
\cite{PRST05} and leptogenesis cannot be successful for $M_1 \ltap
10^{12}~{\rm GeV}$ (i.e. in the regime in which the lepton flavour
effects are significant).  It was suggested in \cite{PPRio106} that
the observed value of $Y_B$ can be reproduced for $M_1 \ltap
10^{12}~{\rm GeV}$ in the case of IH spectrum and real (CP-conserving)
elements of $R$ if the lightest neutrino mass $m_3$ is non-negligible,
having a value in the interval $10^{-2}\sqrt{\deltasol} \ltap m_3
\ltap 0.5\sqrt{\deltasol}$, where $\deltasol = \Delta m^2_{21} \equiv
m^2_2 - m^2_1 \cong 8.0\times 10^{-5}~{\rm eV^2}$ is the mass squared
difference responsible for the solar neutrino oscillations, and
$m_{1,2}$ are the masses of the two additional light Majorana
neutrinos. In this case we still would have $m_3 \ll m_1,m_2$ since
$m_{1,2} \cong \sqrt{|\deltaatm|} \cong 5.0\times 10^{-2}~{\rm eV}$,
$|\deltaatm| \equiv m^2_2 - m^2_3 \cong m^2_1 - m^2_3$ being the mass
squared difference associated with the dominant atmospheric neutrino
oscillations.

 It should be noted that constructing a viable see-saw model 
which leads to real or purely imaginary matrix $R$ might 
encounter serious difficulties, as two recent attemps in 
this direction indicate \cite{antusch,Casas07}. 
However, constructing such a model lies outside the scope of 
our study.

  In the present article we investigate the effects of the lightest
neutrino mass on ``flavoured'' (thermal) leptogenesis. We concentrate
on the case when the CP-violation necessary for the generation of the
observed baryon asymmetry of the Universe is due exclusively to the
Dirac and/or Majorana CP-violating phases in the PMNS matrix $U$.  Our
study is performed within the simplest type I see-saw scenario with
three heavy RH Majorana neutrinos $N_j$, $j=1,2,3$.  The latter are
assumed to have a hierarchical mass spectrum, $M_1 \ll M_{2,3}$.
Throughout the present study we employ the ``orthogonal''
parametrisation of the matrix of neutrino Yukawa couplings
\cite{Casas01}.  As was already mentioned earlier, this
parametrisation involves an orthogonal matrix $R$, $R^TR = RR^T = {\bf
  1}$.  Although, in general, the matrix $R$ can be complex, i.e.
CP-violating, in the present work we are primarily interested in the
possibility that $R$ conserves the CP-symmetry.  We consider the two
types of light neutrino mass spectrum allowed by the data (see, e.g.
\cite{STPNu04}): i) with normal ordering ($\deltaatm >0$), $m_1 < m_2
< m_3$, and ii) with inverted ordering ($\deltaatm < 0$), $m_3 < m_1 <
m_2$.  The case of inverted hierarchical (IH) spectrum and real (and
CP-conserving) matrix $R$ is investigated in detail.  Results for the
normal hierarchical (NH) spectrum are also presented.

Our analysis is performed for negligible renormalisation group (RG)
running of $m_j$ and of the parameters in the PMNS matrix $\pmns$ from
$M_Z$ to $M_1$.  This possibility is realised (in the class of
theories of interest) for sufficiently small values of the lightest
neutrino mass ${\rm min}(m_j)$~\cite{rad,PST06}, e.g., for ${\rm
  min}(m_j) \ltap 0.10$ eV.  The latter condition is fulfilled for the
NH and IH neutrino mass spectra, as well as for spectrum with partial
hierarchy (see, e.g. \cite{BPP1}).  Under the indicated condition
$m_j$, and correspondingly $\deltaatm$ and $\deltasol$, and $U$ can be
taken at the scale $\sim M_Z$, at which the neutrino mixing parameters
are measured.

Throughout the present work we use the standard parametrisation of the
PMNS matrix:
\bea 
\label{eq:Upara}
\pmns = \left( \bad 
c_{12} c_{13} & s_{12} c_{13} & s_{13}e^{-i \delta}  \\[0.2cm] 
 -s_{12} c_{23} - c_{12} s_{23} s_{13} e^{i \delta} 
 & c_{12} c_{23} - s_{12} s_{23} s_{13} e^{i \delta} 
& s_{23} c_{13}  \\[0.2cm] 
 s_{12} s_{23} - c_{12} c_{23} s_{13} e^{i \delta} & 
 - c_{12} s_{23} - s_{12} c_{23} s_{13} e^{i \delta} 
 & c_{23} c_{13} \\ 
                \ea   \right) 
~{\rm diag}(1, e^{i \frac{\alpha_{21}}{2}}, e^{i \frac{\alpha_{31}}{2}})
\eea
%
\noindent where $c_{ij} \equiv \cos\theta_{ij}$, $s_{ij} \equiv
\sin\theta_{ij}$, $\theta_{ij} = [0,\pi/2]$, $\delta = [0,2\pi]$ is
the Dirac CP-violating (CPV) phase and $\alpha_{21}$ and $\alpha_{31}$
are the two Majorana CPV phases \cite{BHP80,SchValle80Doi81},
$\alpha_{21,31} = [0,2\pi]$.  All our numerical results are obtained
for the current best fit values of the solar and atmospheric neutrino
oscillation parameters \cite{BCGPRKL2,TSchwSNOW06,Fogli06},
$\deltasol$, $\sin^2 \theta_{12}$ and $\deltaatm$, $\sin^2
2\theta_{23}$:
\begin{eqnarray}
\label{deltaatmvalues}
\deltasol = \Delta m^2_{21} =
 8.0 \times 10^{-5} \ \eV^2\,,~~~~~~~~\sin^2 \theta_{12} = 0.30\,,\\ [0.25cm]
|\deltaatm| = |\Delta m^2_{31(32)}| = 
2.5 \times 10^{-3} \ \eV^2\,,~~~~
\sin^2 2\theta_{23} = 1\,.
\end{eqnarray}
%
In certain cases the predictions for $|Y_B|$ are very sensitive to the
variation of $\sin^2\theta_{12}$ and $\sin^22\theta_{23}$ within their
95\% C.L.  allowed ranges:
\beq
0.26 \leq \sin^2 \theta_{12} \leq 0.36\,,~
0.36 \leq \sin^2 \theta_{23} \leq 0.64\,,~~95\%~~{\rm C.L.}
\label{th122395}
\eeq
%
We also use the current upper limit on the CHOOZ mixing angle
$\theta_{13}$ \cite{CHOOZPV,BCGPRKL2,TSchwSNOW06}:
\beq
\sin^2\theta_{13} < 0.025~(0.041)\,,~~~95\%~(99.73\%)~{\rm C.L.}
\label{th13}
\eeq
%
%
\section{Baryon Asymmetry from Low Energy CP-Violating Dirac and
  Majorana Phases in $U_{\rm PMNS}$}
%
%
\indent Following \cite{PPRio106} we perform the analysis in the
framework of the simplest type I see-saw scenario.  It includes the
Lagrangian of the Standard Model (SM) with the addition of three heavy
right-handed Majorana neutrinos $N_{j}$ ($j=1,2,3$) with masses $0 <
M_{1} < M_{2} < M_{3}$ and Yukawa couplings $\lambda_{j l}$,
$l=e,\mu,\tau$.  We will work in the basis in which i) the Yukawa
couplings for the charged leptons are flavour-diagonal, and ii) the
Majorana mass term of the RH neutrino fields is also diagonal.  The
heavy Majorana neutrinos are assumed to possess a hierarchical mass
spectrum, $M_1 \ll M_2 \ll M_3$.

In what follows we will use the well-known ``orthogonal
parametrisation`` of the matrix of neutrino Yukawa couplings
\cite{Casas01}:
\begin{equation}
\label{R}
\lambda = \frac{1}{v} \,  \sqrt{M} \, R\, \sqrt{m}\, U^{\dagger}\;,
\end{equation} 
%
\noindent where $R$ is, in general, a complex orthogonal matrix,
$R~R^T = R^T~R = {\bf 1}$, $M$ and $m$ are diagonal matrices formed by
the masses of $N_j$ and of the light Majorana neutrinos $\nu_j$, $M
\equiv {\rm Diag}(M_1,M_2,M_3)$, $m \equiv {\rm Diag}(m_1,m_2,m_3)$,
$M_j > 0$, $m_k \geq 0$, and $v = 174$ GeV is the vacuum expectation
value of the Higgs doublet field.  We shall assume that the matrix $R$
has real and/or purely imaginary elements.

In the case of ``hierarchical'' heavy Majorana neutrinos $N_j$, the
CP-violating asymmetries, relevant for leptogenesis, are generated in
out-of-equilibrium decays of the lightest one, $N_1$.  The asymmetry
in the lepton flavour $l$ (lepton charge $L_l$) is given by
\cite{davidsonetal,niretal,davidsonetal2}:
\begin{eqnarray}
\label{epsa1}
\epsilon_{l}&=& -\frac{3 M_1}{16\pi v^2}\, \frac{{\rm Im}\left(
\sum_{j k} 
m_j^{1/2}\,m_k^{3/2}\, 
U^*_{lj}\, U_{lk}\,
R_{1j }\, R_{1k}\right)}{\sum_i m_i\, \left|R_{1i}\right|^2}\,.
\end{eqnarray}
%
Thus, for real or purely imaginary elements $R_{1j}$ of $R$,
$\epsilon_{e} + \epsilon_{\mu}+ \epsilon_{\tau} = 0$.

There are three possible regimes of generation of the baryon asymmetry
in the leptogenesis scenario \cite{davidsonetal,niretal,davidsonetal2}.  At
temperatures $T \sim M_1 > 10^{12}$ GeV the lepton flavours are
indistinguishable and the one flavour approximation is valid.  The
relevant asymmetry is $\epsilon = \epsilon_{e}+ \epsilon_{\mu}+
\epsilon_{\tau}$ and in the case of interest (real or purely imaginary
CP-conserving $R_{1j}$) no baryon asymmetry is produced.  For
$10^{9}~{\rm GeV} \ltap T \sim M_1\ltap 10^{12}~{\rm GeV}$, the
Boltzmann evolution of the asymmetry $\epsilon_{\tau}$ in the
$\tau-$flavour (lepton charge $L_\tau$ of the Universe) is
distinguishable from the evolution of the $(e + \mu)-$flavour (or
lepton charge $L_e + L_{\mu}$) asymmetry $\epsilon_{e}+
\epsilon_{\mu}$.  This corresponds to the so-called ``two-flavour
regime''\footnote{As was suggested in~\cite{PPRio106} and confirmed
  in the more detailed study~\cite{DiBGRaff06}, in the two-flavour
  regime of leptogenesis the flavour effects are fully developed at
  $M_1 \ltap 5\times 10^{11}$ GeV.}.  At smaller temperatures, $T \sim
M_1 \ltap 10^9$ GeV, the evolution of the $\mu-$flavour (lepton charge
$L_\mu$) and of $\epsilon_{\mu}$ also become distinguishable
(three-flavour regime).  The produced baryon asymmetry is a sum of the
relevant flavour asymmetries, each weighted by the corresponding
efficiency factor accounting for the wash-out processes.

In the two-flavour regime, $10^{9}~{\rm GeV} \ltap T \sim M_1 \ltap
10^{12}$ GeV, the baryon asymmetry\footnote{The expression we give is
  of the baryon asymmetry normalised to the entropy density, see, e.g.
  \cite{PPRio106}.} predicted in the case of interest is given by
\cite{davidsonetal2} (see also \cite{PPRio106}):
\begin{eqnarray}
Y_B &\!\!\cong\!\!&  -\frac{12}{37 g_*}\left(\epsilon_2\, 
\eta\left(\frac{417}{589}\,\widetilde{m}_2\right) 
+ \epsilon_\tau\,\eta\left(\frac{390}{589}
\,\widetilde{m}_\tau\right)\right)\nonumber \\
 &\!\!=\!\!& -\frac{12}{37 g_*}\, \epsilon_\tau \,
\left( 
\eta\left(\frac{390}{589}
\,\widetilde{m}_\tau\right) -
\eta\left(\frac{417}{589}\,\widetilde{m}_2\right) 
 \right) \,,
\label{YB2f}
\end{eqnarray}
%
where the second expression corresponds to real and purely imaginary
$R_{1j}R_{1k}$.  Here $g_* = 217/2$ is the number of relativistic
degrees of freedom, $\epsilon_2 = \epsilon_{e} + \epsilon_{\mu}$,
$\widetilde{m}_2=\widetilde{m}_e+ \widetilde{m}_\mu$, 
$\widetilde{m}_l$ is the ``wash-out mass parameter'' for the asymmetry
in the lepton flavour $l$ \cite{davidsonetal,niretal,davidsonetal2},
\begin{eqnarray}
  \widetilde{m}_l 
  &=& \left|\sum_{k}R_{1k}\, m_k^{1/2}\, U_{lk}^*\right|^2
  \,, ~~~l=e,\mu,\tau\,,
\label{tildmal1}
\end{eqnarray}
%
and $\eta(390\widetilde{m}_\tau/589)\cong
\eta(0.66\widetilde{m}_\tau)$ and $\eta(417\widetilde{m}_2/589) \cong
\eta(0.71\widetilde{m}_2)$ are the efficiency factors for generation
of the asymmetries $\epsilon_{\tau}$ and $\epsilon_2$. The efficiency
factors are well approximated by the expression \cite{davidsonetal2}:
\begin{equation}
\eta\left(X\right) \cong 
 \left(
\frac{8.25\times 10^{-3}\,{\rm eV}}{X}
 + \left(
\frac{X}{2\times 10^{-4}\,{\rm eV}} 
\right)^{1.16}\,
\right)^{-1}\,.
\label{eta1}
\end{equation}
%
 
At $T \sim M_1 \ltap 10^{9}$ GeV, 
the three-flavour regime is realised and~\cite{davidsonetal2} 
\begin{equation}
Y_B \cong -\, \frac{12}{37 g_*}\left(
\epsilon_e\, \eta(\frac{151}{179}\, \widetilde{m}_e)
 + \epsilon_{\mu}\, \eta(\frac{344}{537}\, \widetilde{m}_\mu)
 + \epsilon_{\tau}\, \eta(\frac{344}{537}\, \widetilde{m}_\tau)
\right)\,.
\label{YB3f}
\end{equation}
%

For real or purely imaginary $R_{1j}R_{1k}$ of interest, $j \neq k$,
it proves convenient to cast the asymmetries $\epsilon_{l}$ in the
form \cite{PPRio106}:
\begin{eqnarray}
\label{epsa2}
\epsilon_{l}
= -\, \frac{3M_1}{16\pi v^2} 
\frac{\sum_k \sum_{j>k}\sqrt{m_k m_j}\,
(m_j - m_k)\, \rho_{kj}|R_{1k}R_{1j}|\,
{\rm Im}\,\left(U^*_{lk}\, U_{lj}\right)}
{\sum_i m_i\, |R_{1 i}|^2}\,,~{\rm if}~{\rm Im}\,(R_{1k} R_{1j})=0\,,~
\\
\epsilon_{l} = -\, \frac{3M_1}{16\pi v^2} 
\frac{\sum_k \sum_{j>k}\sqrt{m_k m_j}\,
(m_j + m_k)\, \rho_{kj}|R_{1k} R_{1j}|\,
{\rm Re}\,\left(U^*_{lk} U_{lj}\right)}
{\sum_i m_i\, |R_{1 i}|^2}\,,~{\rm if}~{\rm Re}\,(R_{1k}R_{1j})=0\,.~
\label{epsa3}
\end{eqnarray}
%
where we have used $R_{1j}R_{1k} = \rho_{jk}~|R_{1j} R_{1k}|$ and
$R_{1j}R_{1k} = i\rho_{jk}~|R_{1j} R_{1k}|$, $\rho_{jk} = \pm 1$, $j
\neq k$. Note that real (purely imaginary) $R_{1k}R_{1j}$ and purely
imaginary (real) $U^*_{lk} U_{lj}$, $j\neq k$, implies violation of
CP-invariance by the matrix $R$ \cite{PPRio106}. In order for the
CP-symmetry to be broken at low energies, we should have both ${\rm
  Re}(U^*_{lk} U_{lj})\neq 0$ and ${\rm Im}(U^*_{lk} U_{lj})\neq 0$
(see \cite{PPRio106} for further details).  Note also that if
$R_{1j}$, $j=1,2,3$, is real or purely imaginary, as the condition of
CP-invariance requires \cite{PPRio106}, of the three quantities
$R_{11}R_{12}$, $R_{11}R_{13}$ and $R_{12}R_{13}$, relevant for our
discussion, not more than two can be purely imaginary, i.e.  if, for
instance, $R_{11}R_{12} = i\rho_{12}~|R_{11} R_{12}|$ and
$R_{12}R_{13} = i\rho_{23}~|R_{12} R_{13}|$, then we will have
$R_{11}R_{13} = \rho_{13}~|R_{11} R_{13}|$.

%
\section{Effects of Lightest Neutrino Mass: Real $R_{1j}$}
%
%

We consider next the possible effects the lightest neutrino mass ${\rm
  min}(m_j)$ can have on (thermal) leptogenesis. We will assume that
the latter takes place in the regime in which the lepton flavour
effects are significant and that the CP-violation necessary for the
generation of the baryon asymmetry is provided only by the Majorana or
Dirac phases in the PMNS matrix $U_{\rm PMNS}$.  In the present
Section we analyse the possibility of real elements $R_{1j}$,
$j=1,2,3$, of the matrix $R$.  The study will be performed both for
light neutrino mass spectrum with normal and inverted ordering.  We
begin with the more interesting possibility of spectrum with inverted
ordering (hierarchy).

%
\subsection{Light Neutrino Mass Spectrum with Inverted Ordering}
%
%

The case of inverted hierarchical (IH) neutrino mass spectrum, $m_3
\ll m_1 < m_2$, $m_{1,2} \cong \sqrt{|\deltaatm|}$, is of particular
interest since, as was already mentioned in the Introduction, for real
$R_{1j}$, $j=1,2,3$, IH spectrum and negligible lightest neutrino mass
$m_3 \cong 0$, it is impossible to generate the observed baryon
asymmetry $Y_B \cong 8.6\times 10^{-11}$ in the regime of
``flavoured'' leptogenesis \cite{PPRio106}, i.e. for $M_1 \ltap
10^{12}~{\rm GeV}$, if the only source of CP violation are the
Majorana and/or Dirac phases in the PMNS matrix.  For $m_3 \ll m_1 <
m_2$ and real $R_{1j}$, the terms proportional to $\sqrt{m_3}$ in the
expressions for the asymmetries $\epsilon_{l}$ and wash-out mass
parameters $\widetilde{m}_l$, $l=e,\mu,\tau$, will be negligible if
$m_3 \cong 0$, or if $R_{13} = 0$ and $R_{11},R_{12}\neq 0$, $R_{11}^2
+ R_{12}^2 = 1$.  The main reason for the indicated negative result
lies in the fact that if $m_3 = 0$, or $m_3 \ll m_1 < m_2$ and $R_{13}
= 0$, the lepton asymmetries $\epsilon_{l}$ are suppressed by the
factor $\deltasol/(2\deltaatm) \cong 1.6 \times 10^{-2}$, while
$|R_{11}|,|R_{12}|\leq 1$, and the resulting baryon asymmetry is too
small
\footnote{This suppression is present also in the ``one-flavour''
  regime of $Y_B$ generation, i.e.  in the sum $\epsilon_{e} +
  \epsilon_{\mu} + \epsilon_{\tau}$, when $R_{13} = 0$ and the product
  $R_{11}R_{12}$ has non-trivial real and imaginary parts
  \cite{PRST05}.}.
 
In what follows we will analyse the generation of the baryon asymmetry
$Y_B$ for real $R_{1j}$, $j=1,2,3$, when $m_3$ is non-negligible. We
will assume that $Y_B$ is produced in the two-flavour regime,
$10^9~{\rm GeV}\ltap M_1 \ltap 10^{12}$ GeV.  Under these conditions
the terms $\propto \sqrt{m_3}$ in $\epsilon_{l}$ will be dominant
provided \cite{PPRio106}
\begin{equation}
2\left( \frac{m_3}{\sqrt{\dmsol}} \right)^{\half}
\left(\frac{\dma}{\dmsol} \right)^{\frac{3}{4}}
\frac{\left|R_{13}\right|}{\left|R_{11(12)}\right|}\gg 1\,. 
\label{limit}
\end{equation}
%
This inequality can be fulfilled if $R_{11}\ra 0$, or $R_{12}\ra 0$,
and if $m_3$ is sufficiently large.  The neutrino mass spectrum will
be of the IH type if $m_3$ still obeys $m_3 \ll m_{1,2}$.  The latter
condition can be satisfied for $m_3$ having a value $m_3 \ltap 5\times
10^{-3}~{\rm eV} \ll \sqrt{|\deltaatm|}$.  Our general analysis will
be performed for values of $m_3$ from the interval $10^{-10}~{\rm eV}
\ltap m_3 \ltap 5\times 10^{-2}~{\rm eV}$.

Consider the simple possibility of $R_{11} = 0$.  We will present
later results of a general analysis, performed without setting
$R_{11}$ to 0. For $R_{11} = 0$ the asymmetry $\epsilon_{\tau} = -
(\epsilon_{e} + \epsilon_{\mu})$ of interest is given by:
\begin{equation}
\epsilon_\tau \, \cong \;
-\, \frac{3M_1}{16\pi v^2}\sqrt{m_3\, m_2}\, 
\left (1 - \frac{m_3}{m_2} \right )\, \rho_{23}\, r\,
{\rm Im}\left(U^{*}_{\tau 2}U_{\tau 3}\right)\,,
\label{IHepst1}
\end{equation}
%
where
\begin{eqnarray}
m_2 = \sqrt{m^2_3 + |\deltaatm|}\,, \\[0.25cm]
r\;=\;\frac{\left|R_{13} R_{12}\right|}
{|R_{12}|^2+\frac{m_3}{m_2} |R_{13}|^2}\,,
\label{IHr1}
\end{eqnarray}
%
and
\begin{equation}
{\rm Im}\left(U^{*}_{\tau 2}U_{\tau 3}\right)\;=\;
-\, c_{23}c_{13}\, {\rm Im}\left(e^{i(\alpha_{31}-\alpha_{21})/2}
(c_{12}s_{23} + s_{12}c_{23}s_{13}e^{-i\delta})\right)\,.
\label{IHIm23}
\end{equation}
%
The two relevant wash-out mass parameters are given by:
\begin{eqnarray} 
\label{tmtau1}
\widetilde{m}_\tau &=& 
m_2\, R_{12}^2|U_{\tau 2}|^2 + m_3\, R_{13}^2 |U_{\tau 3}|^2 + 
2\,\sqrt{m_2\, m_3}\,\, \rho_{23}\,|R_{12}R_{13}|\,
{\rm Re}\,\left (U^*_{\tau 2}U_{\tau 3}\right ),~~~~~~
\\ [0.25cm]
\widetilde{m}_2 &\equiv& \widetilde{m}_e + \widetilde{m}_\mu
= m_2\, R_{12}^2 + m_3\, R_{13}^2 - \widetilde{m}_\tau\,,
\label{IHtm1}
\end{eqnarray}
%
where $\rho_{23} = {\rm sgn}(R_{12}R_{13})$. 

The orthogonality of the matrix $R$ implies that $R_{11}^2 + R_{12}^2
+ R_{13}^2 = 1$, which in the case under consideration reduces to
$R_{12}^2 + R_{13}^2 = 1$.  It is not difficult to show that for
$R_{12}$ and $R_{13}$ satisfying this constraint, the maximum of the
function $r$, and therefore of the asymmetry $|\epsilon_\tau|$, takes
place for
\begin{equation}
R^2_{12} = \frac{m_3}{m_3 + m_2}\,,~~ 
R^2_{13} = \frac{m_2}{m_3 + m_2}\,,~~~R^2_{12} < R^2_{13}\,. 
\label{IHmaxrR12}
\end{equation}
%
At the maximum we get
\begin{equation}
{\rm max}(|r|) = \frac{1}{2}\,
\left (\frac{m_2}{m_3} \right )^{\frac{1}{2}} \cong 
\frac{1}{2}\, \left (\frac{\sqrt{|\deltaatm|}}{m_3} \right )^{\frac{1}{2}}
\,,
\label{IHmaxr1}
\end{equation}
%
and
\begin{equation}
|\epsilon_\tau | \cong 
\frac{3M_1}{32\pi v^2}\, \left ( m_2 - m_3 \right )\, 
\left | {\rm Im}\left(U^{*}_{\tau 2}U_{\tau 3}\right) \right | 
\cong \frac{3M_1}{32\pi v^2}\sqrt{|\deltaatm|}\, 
\left | {\rm Im}\left(U^{*}_{\tau 2}U_{\tau 3}\right) \right |\,. 
\label{IOmaxepst1}
\end{equation}
%
The second approximate equalities in eqs.~(\ref{IHmaxr1}) and
(\ref{IOmaxepst1}) correspond to IH spectrum, i.e. to $m_3 \ll m_2
\cong \sqrt{|\deltaatm|}$.  Thus, the maximum of the asymmetry
$|\epsilon_\tau|$ thus found i) is not suppressed by the factor
$\deltasol/(\deltaatm)$, and ii) practically does not depend on $m_3$
in the case of IH spectrum. Given the fact that
\begin{equation}
  |\epsilon_\tau | \cong 
  5.0\times 10^{-8}\, \frac{m_2 - m_3}{\sqrt{|\deltaatm|}}\, 
  \left (\frac{\sqrt{|\deltaatm|}}{0.05~{\rm eV}}\right )
  \left (\frac{M_1}{10^{9}~{\rm GeV}}\right )
  \left | {\rm Im}\left(U^{*}_{\tau 2}U_{\tau 3}\right) \right |~\,,
\label{IOmaxepst2}
\end{equation}
%
${\rm max}(|{\rm Im}(U^{*}_{\tau 2}U_{\tau 3})|) \cong 0.46$, where we
have used $\sin^22\theta_{23} = 1$, $\sin^2\theta_{12} = 0.30$ and
$\sin^2\theta_{13} < 0.04$, and that ${\rm
  max}(|\eta(0.66\widetilde{m}_\tau) - \eta(0.71\widetilde{m}_2)|)
\cong 7\times 10^{-2}$, we find the absolute upper bound on the baryon
asymmetry in the case of IH spectrum and real matrix $R$ (real
$R_{1j}R_{1k}$):
\begin{equation}
|Y_B | \ltap 4.8\times 10^{-12}
\left (\frac{\sqrt{|\deltaatm|}}{0.05~{\rm eV}}\right )
\left (\frac{M_1}{10^{9}~{\rm GeV}}\right )~.
\label{IHmaxYB1}
\end{equation}
%
This upper bound allows to determine the minimal value of $M_1$ for
which it is possible to reproduce the observed value of $|Y_B |$ lying
in the interval $ 8.0\times 10^{-11} \ltap |Y_B | \ltap 9.2\times
10^{-11}$ for IH spectrum, real $R$ and $R_{11} = 0$:
\begin{equation}
M_1 \gtap 1.7\times 10^{10}~{\rm GeV}~.
\label{IHminM1}
\end{equation}
%

The values of $R_{12}$, for which $|\epsilon_\tau|$ is maximal, can
differ, in general, from those maximising $|Y_B|$ due to the
dependence of the wash-out mass parameters and of the corresponding
efficiency factors on $R_{12}$.  However, this difference, when it is
present, does not exceed 30\%, as our calculations show, and is not
significant. At the same time the discussion of the wash-out effects
for the maximal $|\epsilon_\tau|$ is rather straightforward and allows
to understand in a rather simple way the specific features of the
generation of $|Y_B|$ in the case under discussion.  For these reasons
in our discussion of the wash-out mass parameters we will use $R_{12}$
maximising $|\epsilon_\tau|$.  All our major numerical results and
most of the figures are obtained for $R_{12}$ maximising $|Y_B|$.

For $R_{12}$ ($R_{13}$), which maximises the ratio $|r|$ and the
asymmetry $|\epsilon_{\tau}|$, the relevant wash-out mass parameters
are given by:
\begin{eqnarray} 
\label{IOtmtau1}
\widetilde{m}_\tau &=& 
\frac{m_2\,m_3}{m_3 + m_2}\,
\left [~ 
|U_{\tau 2}|^2 +  |U_{\tau 3}|^2 + 
2\rho_{23} {\rm Re}\,\left (U^*_{\tau 2}U_{\tau 3}\right )
\right ]\,, \\ [0.25cm]
\widetilde{m}_2 &=& 2\, 
\frac{m_2\,m_3}{m_3 + m_2}  - \widetilde{m}_\tau\,.
\label{IOtm21}
\end{eqnarray}
%

Equations (\ref{IOmaxepst2}), (\ref{IOtmtau1}) and (\ref{IOtm21})
suggest that in the case of IH spectrum with non-negligible $m_3$,
$m_3 \ll \sqrt{|\deltaatm|}$, the generated baryon asymmetry $|Y_B|$
can be strongly enhanced in comparison with the asymmetry $|Y_B|$
produced if $m_3 \cong 0$. The enhancement can be by a factor of $\sim
100$. Indeed, the maximum of the asymmetry $|\epsilon_\tau|$ (with
respect to $|R_{12}|$), eq.~(\ref{IOmaxepst1}), does not contain the
suppression factor $\deltasol/(2\deltaatm) \cong 1.6\times 10^{-2}$
and its magnitude is not controlled by $m_3$, but rather by
$\sqrt{|\deltaatm|}$.  At the same time, the wash-out mass parameters
$\widetilde{m}_\tau$ and $\widetilde{m}_2$, eqs.~(\ref{IOtmtau1}) and
(\ref{IOtm21}), are determined by $m_2m_3/(m_2+m_3) \cong m_3$.  The
latter in the case under discussion can take values as large as $m_3
\sim 5\times 10^{-3}$ eV.  The efficiency factors
$\eta(0.66\widetilde{m}_\tau)$ and $\eta(0.71\widetilde{m}_2)$, which
enter into the expression for the baryon asymmetry, eq.~(\ref{YB2f}),
have a maximal value $\eta(X) \cong (6 - 7)\times 10^{-2}$ when $X
\cong (0.7 - 1.5)\times 10^{-3}$ eV (weak wash-out regime).  Given the
range of values of $m_3$ for IH spectrum extends to $\sim 5\times
10^{-3}$ eV, one can always find a value of $m_3$ in this range such
that $\widetilde{m}_\tau$ or $\widetilde{m}_2$ take a value maximising
$\eta(0.66\widetilde{m}_\tau)$ or $\eta(0.71\widetilde{m}_2)$, and
$|\eta(0.66\widetilde{m}_\tau) - \eta(0.71\widetilde{m}_2)|$.  This
qualitative discussion suggests that there always exists an interval
of values of $m_3$ for which the baryon asymmetry is produced in the
weak wash-out regime.  On the basis of the above considerations one
can expect that we can have successful leptogenesis for a
non-negligible $m_3$ in the case of IH spectrum even if the requisite
CP-violation is provided by the Majorana or Dirac phase(s) in the PMNS
matrix. This is confirmed by the detailed (analytic and numerical)
analysis we have performed,
the results of which are described below.\\

{\bf {\it A. Leptogenesis due to Majorana 
CP-Violation in $U_{\rm PMNS}$ 
}}\\

We will assume first that the Dirac phase $\delta$ has a CP-conserving
value, $\delta =0;~\pi$.  For $\delta = 0~(\pi)$, we have $|{\rm
  Im}(U^*_{\tau 2}U_{\tau 3})| = c_{23}c_{13}(s_{23}c_{12}~
^{~+}_{(-)}~ c_{23}s_{12}s_{13})|\sin\alpha_{32}/2|$ and
correspondingly $0.36|\sin\alpha_{32}/2|\ltap |{\rm Im}(U^*_{\tau
  2}U_{\tau 3})| \ltap 0.46|\sin\alpha_{32}/2|$, where $\alpha_{32} =
\alpha_{31} - \alpha_{21}$ and we have used the best fit values of
$\sin^22\theta_{23}$ and $\sin^2\theta_{12}$, and the limit
$\sin^2\theta_{13} < 0.04$.  For $s_{13} = 0$ we get: $|{\rm
  Im}(U^*_{\tau 2}U_{\tau 3})| \cong 0.42 |\sin\alpha_{32}/2|$.  The
terms proportional to $s_{13}$ have a subdominant effect on the
magnitude of the calculated $|\epsilon_\tau|$ and $|Y_B|$.

It is easy to check that the asymmetry $|\epsilon_\tau|$ and the
wash-out mass parameters $\widetilde{m}_{\tau,2}$ remain invariant
with respect to the change $\rho_{23} \rightarrow - \rho_{23}$,
$\alpha_{32} \rightarrow 2\pi - \alpha_{32}$.  Thus, the baryon
asymmetry $|Y_B|$ satisfies the following relation:
$|Y_B(\rho_{23},\alpha_{32})| = |Y_B(-\rho_{23},2\pi - \alpha_{32})|$.
Therefore, unless otherwise stated, we will consider the case of
$\rho_{23} = + 1$ in what follows.

The absolute maximum of the asymmetry $|Y_B|$ with respect to
$\alpha_{32}$ is not obtained for $\alpha_{32} = \pi$ for which
$|\epsilon_\tau|$ is maximal\footnote{We would like to recall that in
  the case of $\alpha_{32} = \pi$, $\delta = 0;~\pi$, and real
  $R_{1j}R_{1k}$, the requisite violation of CP-symmetry in
  leptogenesis is provided by the matrix $R$ \cite{PPRio106}.}, 
but rather for $\alpha_{32}$ having a value in the interval $\alpha_{32}
\cong (\pi/2 - 2\pi/3)$ if $\rho_{23} = + 1$, or in the interval
$\alpha_{32} \cong (4\pi/3- 3\pi/2)$ if $\rho_{23} = - 1$.  The
maximal value of $|Y_B|$ at $\alpha_{32} = \pi$ is smaller at least by
a factor of $\sim 2$ than the value of $|Y_B|$ at its absolute maximum
(see further Fig.~3). As can be easily shown, for $\alpha_{32} \sim
\pi$ there is a rather strong mutual compensation between the
asymmetries in the lepton charges $L_{\tau}$ and $(L_{e} + L_{\mu})$
owing to the fact that, due to ${\rm Re}(U^*_{\tau 2}U_{\tau 3})=0$,
$\widetilde{m}_\tau$ and $\widetilde{m}_2$ have relatively close
values and $|\eta(0.66\widetilde{m}_\tau) -
\eta(0.71\widetilde{m}_2)|\ltap 10^{-2}$.  Actually, in certain cases
one can even have $|\eta(0.66\widetilde{m}_\tau) -
\eta(0.71\widetilde{m}_2)|\cong 0$, and thus $|Y_B|\cong 0$, for
$\alpha_{32}$ lying in the interval $\alpha_{32} \sim (\pi - 4\pi/3)$
(see further Fig.~3). Similar cancellation can occur for $s_{13} = 0.2$ at
$\alpha_{32} \sim \pi/6$.
Obviously, we have $|Y_B|= 0$ for $\alpha_{32} =0;~2\pi$.

We are interested primarily in the dependence of $|Y_B|$ on $m_3$.  As
$m_3$ increases from the value of $10^{-10}$ eV up to $10^{-4}$ eV, in
the case of $R_{11} = 0$ under discussion the maximal possible $|Y_B|$
for a given $M_1$ increases monotonically, starting from a value which
for $M_1 \leq 10^{12}$ GeV is much smaller than the observed one,
${\rm max}(|Y_B|) \ll 8.6\times 10^{-11}$.  At approximately $m_3
\cong 2\times 10^{-6}$ eV, we have ${\rm max}(|Y_B|) \cong 8.6\times
10^{-11}$ for $M_1 \cong 5\times 10^{11}$ GeV.  As $m_3$ increases
beyond $2\times 10^{-6}$ eV, ${\rm max}(|Y_B|)$ for a given $M_1$
continues to increase until it reaches a maximum. This maximum occurs
for $m_3$ such that $0.71\widetilde{m}_2 \cong 9.0\times 10^{-4}$ eV
and $\eta(0.71\widetilde{m}_2)$ is maximal,
$\eta(0.71\widetilde{m}_{2}) \cong 6.8\times 10^{-2}$, while
$\eta(0.66\widetilde{m}_{\tau})$ is considerably smaller.  As can be
shown, for $\rho_{23} = +1$, it always takes place at $\alpha_{32}
\cong \pi/2$.  For $\alpha_{32} = \pi/2$, $s_{13} = 0$ and $\rho_{23}
= + 1$, the maximum of $|Y_B|$ in question is located at $m_3 \cong
7\times 10^{-4}$ eV.  It corresponds to the CP-asymmetry being
predominantly in the $(e +\mu)-$flavour.  As $m_3$ increases further,
$|\eta(0.66\widetilde{m}_{\tau}) - \eta(0.71\widetilde{m}_{2})|$ and
correspondingly $|Y_B|$, rapidly decrease.  At certain value of $m_3$,
typically lying in the interval $m_3 \sim (1.5 - 2.5)\times 10^{-3}$
eV, one has
\begin{figure}[t!!]
\begin{center}
\vspace{-1cm}
\includegraphics[width=13.5cm,height=9.5cm]{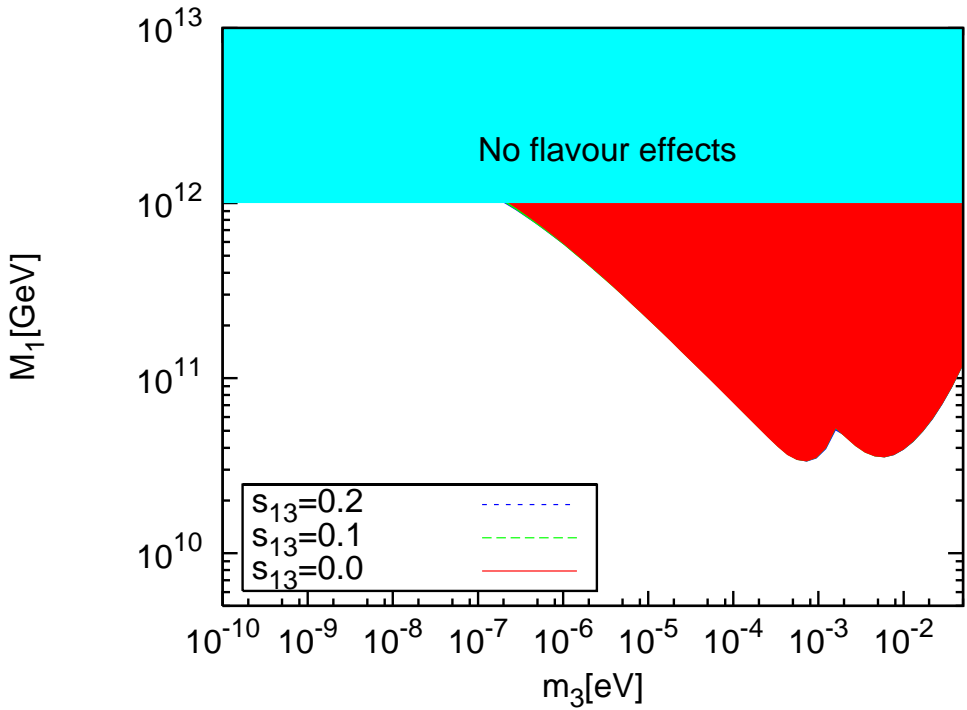}
\caption{
  \label{nfig1IOM1m3.eps} Values of $m_3$ and $M_1$ for which the
  ``flavoured'' leptogenesis is successful, generating baryon
  asymmetry $|Y_B| = 8.6\times 10^{-11}$ (red/dark shaded area). The
  figure corresponds to hierarchical heavy Majorana neutrinos, light
  neutrino mass spectrum with inverted ordering (hierarchy), $m_3 <
  m_1 < m_2$, and real elements $R_{1j}$ of the matrix $R$.  The
  minimal value of $M_1$ at given $m_3$, for which the measured value
  of $|Y_B|$ is reproduced, corresponds to CP-violation due to the
  Majorana phases in the PMNS matrix.  The results shown are obtained
  using the best fit values of neutrino oscillation parameters:
  $\deltasol = 8.0\times 10^{-5}~{\rm eV}^2$, $\deltaatm = 2.5\times
  10^{-3}~{\rm eV}^2$, $\sin^2\theta_{12}=0.30$ and
  $\sin^22\theta_{23}=1$.  See text for further details.}
\end{center}
\end{figure}
%
\noindent $|\eta(0.66\widetilde{m}_{\tau}) -
\eta(0.71\widetilde{m}_{2})| \cong 0$ and $|Y_B|$ goes through a deep
minimum: one can have even $|Y_B| = 0$. This minimum of $|Y_B|$
corresponds to a partial or complete cancellation between the
asymmetries in the $\tau-$flavour and in the $(e
+\mu)-$flavour\footnote{A general discussion of the possibility of
  such a suppression of the baryon asymmetry $Y_B$ in the case of
  ``flavoured'' leptogenesis is given in \cite{PPRio106}.}.  
In our
example of $\alpha_{32} = \pi/2$, $s_{13} = 0$ and $\rho_{23} = + 1$,
the indicated minimum of $|Y_B|$ occurs at $m_3 \cong 2.3\times
10^{-3}$ eV.  As $m_3$ increases further,
$|\eta(0.66\widetilde{m}_{\tau}) - \eta(0.71\widetilde{m}_{2})|$ and
$|Y_B|$ rapidly increase and $|Y_B|$ reaches a second maximum, which
in magnitude is of the order of the first one.  This maximum
corresponds to the CP-asymmetry being predominantly in the $\tau-$
flavour rather than in the $(e +\mu)-$flavour: now
$\eta(0.66\widetilde{m}_{\tau})$ is maximal having a value
$\eta(0.66\widetilde{m}_{\tau}) \cong 6.8\times 10^{-2}$, and
$\eta(0.71\widetilde{m}_{2})$ is substantially smaller.  For
$\rho_{23} = +1$, $s_{13} = 0$ or $s_{13} = 0.2$ and $\delta = 0$, it
takes place at a value of $\alpha_{32}$ close to $\pi/2$, $\alpha_{32}
\cong \pi/2$, while for $s_{13} = 0.2$ and $\delta = \pi$, it occurs
at $\alpha_{32} \cong 2\pi/3$.  In the case of $\rho_{23} = + 1$,
$s_{13} = 0$ and $\alpha_{32} = \pi/2$, the second maximum of $|Y_B|$
is located at $m_3 \cong 7\times 10^{-3}$ eV.  As $m_3$ increases
further, $|Y_B|$ decreases rather slowly monotonically.

These features of the dependence of $|Y_B|$ on $m_3$ are confirmed by
a more general analysis in which, in particular, the value of $R_{11}$
was not set to zero a priori.  The results of this analysis are
presented in Fig.~1, while Fig.~2 illustrates the dependence of
$|Y_B|$ on $m_3$ in the case of $R_{11} = 0$.

In Fig.~1 we show the correlated values of $M_1$ and $m_3$ for which
one can have successful leptogenesis in the case of neutrino mass
spectrum with inverted ordering and CP-violation due to the Majorana
and Dirac phases in $U_{\rm PMNS}$.  The figure was obtained by
performing, for given $m_3$ from the interval $10^{-10} \leq m_3 \leq
0.05$ eV, a thorough scan of the relevant parameter space searching
for possible enhancement or suppression of the baryon asymmetry with
respect to that found for $m_3 = 0$.  The real elements of the
$R-$matrix of interest, $R_{1j}$, $j=1,2,3$, were allowed to vary in
their full ranges determined by the condition of orthogonality of the
matrix $R$: $R^2_{11} + R^2_{12} + R^2_{13} = 1$.  In particular, {\it
  $R_{11}$ was not set to zero}. The Majorana phases $\alpha_{21,31}$
were varied in the interval $[0,2\pi]$.  The calculations were
performed for three values of the CHOOZ angle $\theta_{13}$,
corresponding to $\sin\theta_{13} = 0;~0.1;~0.2$.  In the cases of
$\sin\theta_{13}\neq 0$, the Dirac phase $\delta$ was allowed to take
values in the interval $[0,2\pi]$. The heavy Majorana neutrino mass
$M_1$ was varied in the interval $10^{9}~{\rm GeV} \leq M_1 \leq
10^{12}$ GeV.  For given $m_3$, the minimal value of the mass $M_1$,
for which the leptogenesis is successful, generating $|Y_B| \cong
8.6\times 10^{-11}$, was obtained for the values of the other
parameters which maximise $|Y_B|$. The ${\rm min}(M_1)$ thus found
does not exhibit any significant dependence on $s_{13}$.  If $m_3
\ltap 2.5\times 10^{-7}$ eV, leptogenesis cannot be successful for
$M_1 \leq 10^{12}$ GeV: the baryon asymmetry produced in this regime
is too small. As $m_3$ increases starting from the indicated value,
the maximal $|Y_B|$ for a given $M_1 \leq 10^{12}$ GeV, increases
monotonically. Correspondingly, the ${\rm min}(M_1)$ for which one 
can
have successful leptogenesis decreases monotonically and for 
$m_3 \gtap 5\times 10^{-6}$ eV we have ${\rm min}(M_1) \ltap 5\times
10^{11}$ GeV.  The first maximum of $|Y_B|$ (minimum of $M_1$) as
$m_3$ increases is reached at $m_3 \cong 5.5\times 10^{-4}$ eV,
$\alpha_{32} \cong \pi/2$ ($\alpha_{21} \cong 0.041$, $\alpha_{31}
\cong 1.65$), $R_{11} \cong -0.061$, $R_{12} \cong 0.099$, and $R_{13}
\cong 0.99$.  At the maximum we have $|Y_B| = 8.6\times 10^{-11}$ for
$M_1 \cong 3.4\times 10^{10}$ GeV.  The second maximum of $|Y_B|$ (or
minimum of $M_1$) seen in Fig.~1 corresponds to $m_3 \cong 5.9\times
10^{-3}$ eV, $\alpha_{32} \cong \pi/2$ ($\alpha_{21} \cong -0.022$,
$\alpha_{31} \cong 1.45$), $R_{11} \cong -0.18$, $R_{12} \cong 0.29$,
and $R_{13} \cong -0.94$.  The observed value of $|Y_B|$ is reproduced
in this case for $M_1 \cong 3.5\times 10^{10}$ GeV.

   Similar features are seen in Fig.~2, which shows the dependence of
$|Y_B|$ on $m_3$ for $R_{11} = 0$, fixed $M_1 = 10^{11}$ GeV,
$\alpha_{32} = \pi/2$ , $s_{13} = 0$ and $\rho_{23} = + 1;~(-1)$.  In
the case of $\alpha_{32} = \pi/2$, $s_{13} = 0.2$, $\delta=0$ and
$\rho_{23} = + 1$, the absolute maximum of $|Y_B|$ is obtained for
$m_3 \cong 6.7\times 10^{-3}$ eV and $|R_{12}| = 0.34$ (Fig.~3).  At
this maximum we have $\eta(0.66\widetilde{m}_{\tau}) \cong 0.067$,
\begin{figure}[t!!]
\begin{center}
\vspace{-1.0cm}
\begin{tabular}{cc}
\includegraphics[width=8truecm,height=6.5cm]{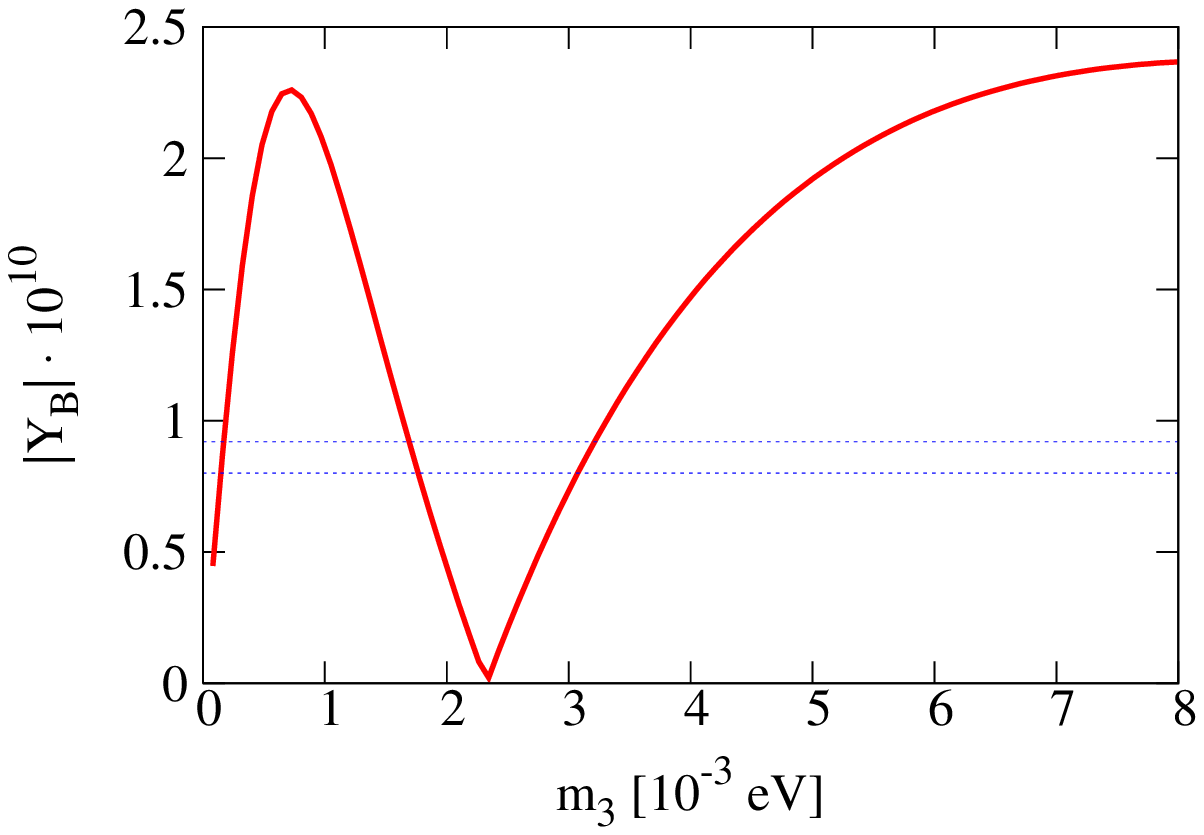}&
\includegraphics[width=8truecm,height=6.5cm]{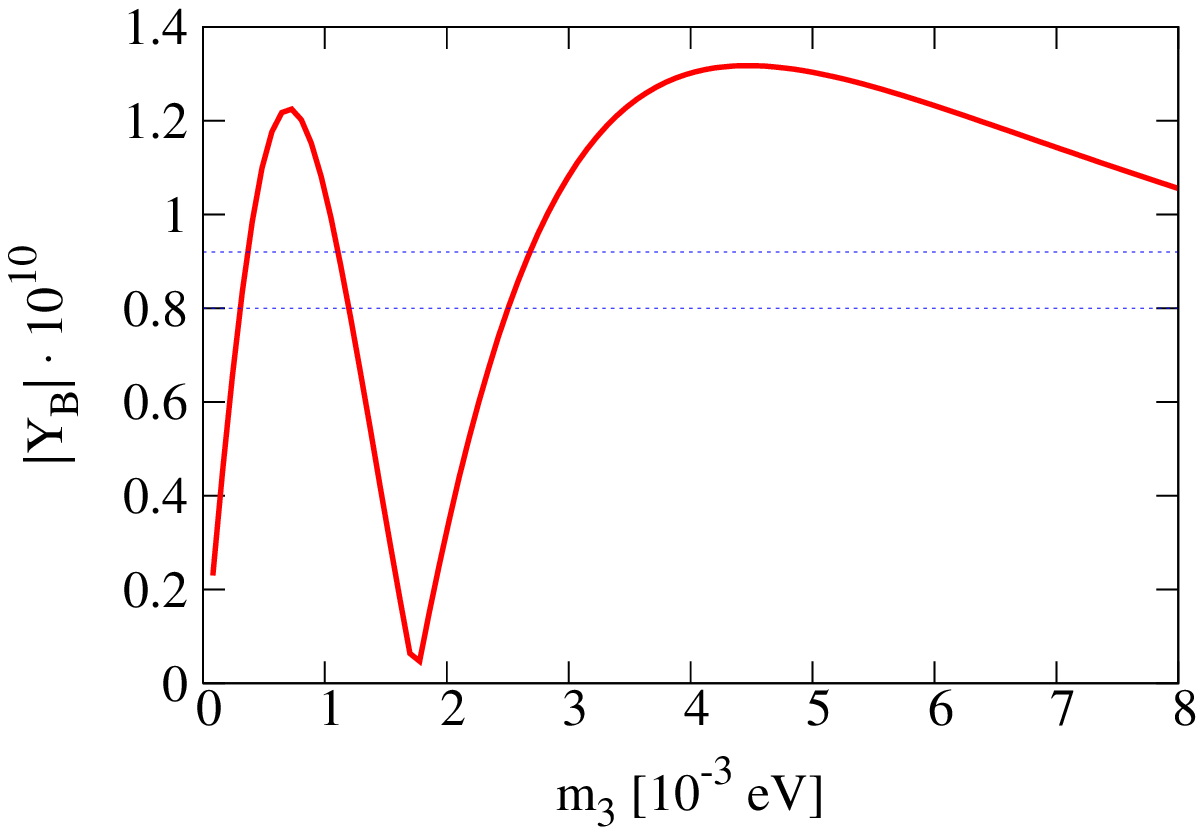}
\end{tabular}
\caption{\label{nEMFig12YBm3.eps} The dependence of $|Y_B|$ on $m_{3}$ in
  the case of IH spectrum, real $R_{1j}R_{1k}$, Majorana CP-violation,
  $R_{11}=0$, $\alpha_{32} = \pi/2$, $s_{13}=0$, $M_1=10^{11}~{\rm
    GeV}$, and for i) ${\rm sgn}(R_{12}R_{13})=+1$ (left panel), and
  ii) ${\rm sgn}(R_{12}R_{13})=-1$ (right panel).  The baryon
  asymmetry $|Y_B|$ was calculated for a given $m_3$, using the value
  of $|R_{12}|$, for which the asymmetry $|\epsilon_\tau |$ is
  maximal.  The horizontal dotted lines indicate the allowed range of
  $|Y_B|$, $|Y_B| = (8.0 - 9.2)\times 10^{-11}$.  }
\end{center}
\end{figure}
%
$\eta(0.71\widetilde{m}_{2})\cong 0.013$, and 
\begin{equation}
|Y_{B} | \cong  2.6\times 10^{-12}\,
\left (\frac{\sqrt{|\deltaatm|}}{0.05~{\rm eV}}\right )\,
\left (\frac{M_1}{10^{9}~{\rm GeV}}\right )\,.
\label{IHmaxYBM1}
\end{equation}
%
Correspondingly, the observed baryon asymmetry $|Y_B|$, $8.0\times
10^{-11}\ltap |Y_B|\ltap 9.2\times 10^{-11}$, can be reproduced if
$M_1\gtap 3.0\times 10^{10}{\rm \;GeV}$.  If $s_{13} = 0$, the same
result holds for $M_1\gtap 3.5 \times 10^{10}{\rm \;GeV}$.  The
minimal values of $M_1$ thus found are somewhat smaller than ${\rm
  min}(M_1)\cong 5.3 \times 10^{10}{\rm \;GeV}$ obtained in the case
of negligible $m_3 \cong 0$ ($R_{13} = 0$) and purely imaginary
$R_{11}R_{12}$ \cite{PPRio106}.  The dependence of the baryon
asymmetry on $\alpha_{32}$ in the case of $s_{13} = 0;~0.2$ discussed
above is illustrated in Fig.~3.
\begin{figure}[t!!]
\begin{center}
\vspace{-1.0cm}
\begin{tabular}{cc}
\includegraphics[width=8truecm,height=6.5cm]{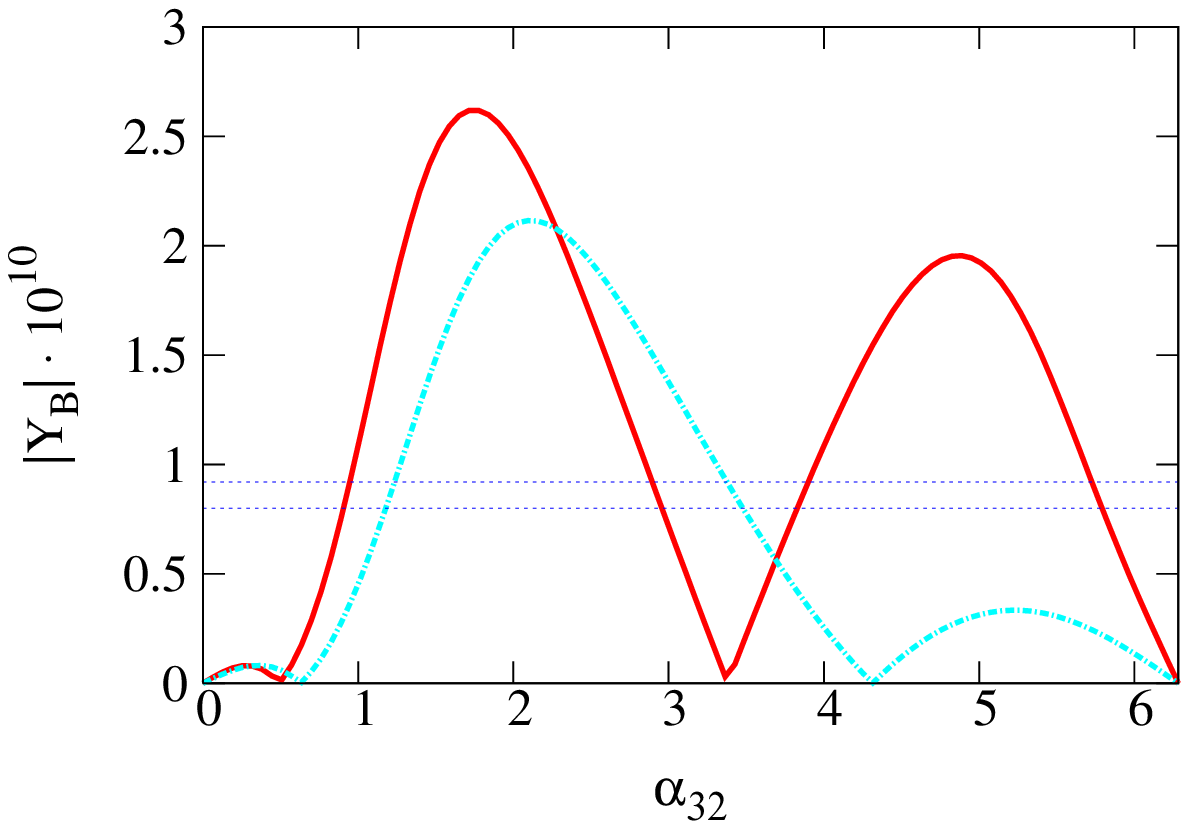}&
\includegraphics[width=8truecm,height=6.5cm]{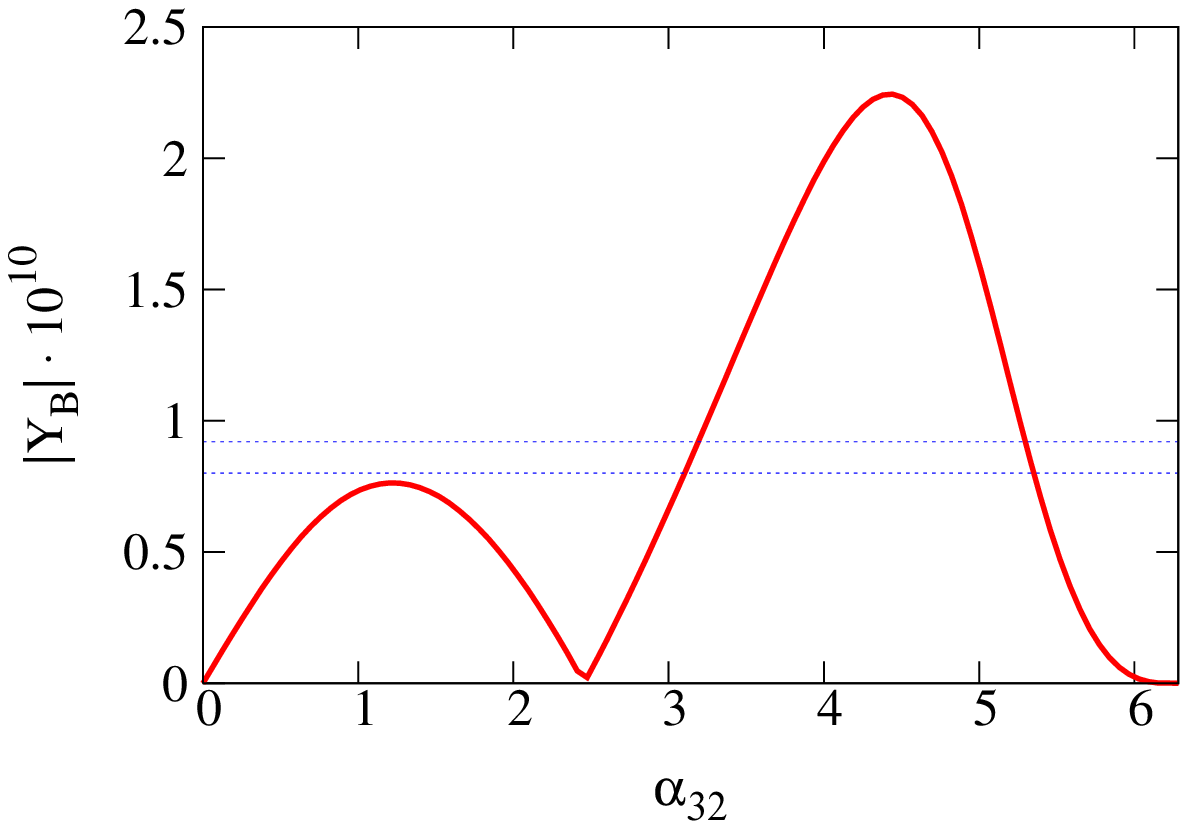}
\end{tabular}
\caption{ \label{nEMfig34s13.eps} The dependence of $|Y_B|$ on
  $\alpha_{32}$ (Majorana CP-violation), in the case of IH spectrum,
  real $R_{1j}R_{1k}$, $R_{11}=0$, $M_1=10^{11}~{\rm GeV}$, and for i)
  $s_{13}=0.2$, $\delta=0~(\pi)$, $|R_{12}|=0.34~(0.38)$, $m_3=
  6.7~(4.3)\times 10^{-3}{\rm eV}$, ${\rm sgn}(R_{12}R_{13})=+1$ (left
  panel, red (blue) line), and ii) $s_{13}=0$, ${\rm
    sgn}(R_{12}R_{13})=-1$, $|R_{12}|=0.41$, $m_3=4.2\times
  10^{-3}{\rm eV}$ (right panel).  The values of $m_3$ and $|R_{12}|$
  used maximise $|Y_B|$ at i) $\alpha_{32} = \pi/2~(2\pi/3)$ and ii)
  $\alpha_{32} = 3\pi/2$.  The horizontal dotted lines indicate the
  allowed range of $|Y_B| = (8.0 - 9.2)\times 10^{-11}$.  }
\end{center}
\end{figure}
%

One obtains similar results in the case of $R_{12} =
0$, $R_{11},R_{13} \neq 0$, we have also analysed. The corresponding
formulae can be obtained from those derived for $R_{11} = 0$ by
replacing $R_{12}$ with $R_{11}$, $U^{*}_{\tau 2}$ with $U^{*}_{\tau
  1}$ and $m_2$ with $m_1 = \sqrt{m^2_3 + |\deltaatm| -\deltasol}
\cong \sqrt{m^2_3 + |\deltaatm|} = m_2$.  In this case we have, in
particular, $|\epsilon_\tau| \propto |{\rm Im}(U^{*}_{\tau 1}U_{\tau
  3})|=$ $|c_{23}c_{13}(s_{12}s_{23} \mp
c_{12}c_{23}s_{13})\sin\alpha_{31}/2|$, where the minus (plus) sign
corresponds to $\delta = 0~(\pi)$.  Evidently, the relevant Majorana
phase\footnote{Note that the Majorana phase $\alpha_{32}$ ($R_{11} =
  0$) or $\alpha_{31}$ ($R_{12} = 0$), relevant for leptogenesis in
  the case of IH spectrum and real matrix $R$, does not coincide with
  the Majorana phase $\alpha_{21}$, which together with
  $\sqrt{|\deltaatm|}$ and $\sin^2\theta_{12}$ determines the values
  of the effective Majorana mass in neutrinoless double beta decay
  (see, e.g.  \cite{BiPet87,BPP1,STPFocusNu04}).} 
is $\alpha_{31}/2$.
Numerically we get $0.19|\sin\alpha_{31}/2| \ltap |{\rm
  Im}(U^{*}_{\tau 1}U_{\tau 3})| \ltap 0.35|\sin\alpha_{31}/2|$, while
for $s_{13} = 0$ we find $|{\rm Im}(U^{*}_{\tau 1}U_{\tau 3})| \cong
0.27|\sin\alpha_{31}/2|$.  Thus, the maximal value of
$|\epsilon_\tau|$ for $R_{12} = 0$ is smaller approximately by a
factor of 1.3 than the maximal value of $|\epsilon_\tau|$ when $R_{11}
= 0$.  As a consequence, the minimal $M_1$ for which we can have
successful leptogenesis can be expected to be bigger by a factor of
$\sim 1.3$ than the one we have obtained in the case of $R_{11} = 0$.
This is confirmed by the numerical calculations we have performed. For
example, for $s_{13}=0.2$, $\delta=\pi$, ${\rm sgn}(R_{12}R_{13})=-1$,
and values of $|R_{11}|= 0.38$ and $m_3= 4.5\times 10^{-3}{\rm eV}$
(which maximise $|Y_B|$ at $\alpha_{31} = 2\pi/3$), we get
\begin{equation}
{\rm max}(|Y_{B}|) \cong  2.2\times 10^{-12}\,
\left (\frac{\sqrt{|\deltaatm|}}{0.05~{\rm eV}}\right )\,
\left (\frac{M_1}{10^{9}~{\rm GeV}}\right )\,.
\label{IHmaxYBM2}
\end{equation}
%
Therefore the observed value of $|Y_{B}|$ can be reproduced for
$M_1 \gtap 3.7\times 10^{10}~{\rm GeV}$.

 We would like to stress that the results we have obtained 
in the case of $R_{1j}\neq 0$, $j=1,2,3$, 
which are shown in Fig.~1, are very different from the results 
for, e.g. $R_{11} = 0$ and $R_{12},R_{13} \neq 0$. 
Along the line of minimal values of $M_1$ 
in Fig.~1, for which we can 
have successful leptogenesis,
we find that   
either $\widetilde{m}_{2} \sim 10^{-3}$ eV and 
$\widetilde{m}_{\tau} \sim 2\times 10^{-4}$ eV, 
or $\widetilde{m}_{\tau} \sim 2\times 10^{-3}$ eV 
and $\widetilde{m}_{2} \gg 10^{-3}$ eV,
practically for any $m_3$ from the 
interval\footnote{As our numerical calculations show, 
at $m_3 \gtap 10^{-9}$ eV, for instance,
there is a very narrow interval of values 
of $m_3$ around  $m_3 \sim 2\times 10^{-3}$ eV,
in which both $\widetilde{m}_{2}$  and 
$\widetilde{m}_{\tau}$ increase rapidly 
monotonically from 
$\sim 10^{-3}$ eV and $\sim 2\times 10^{-4}$ eV to 
$\sim 7\times 10^{-3}$ eV and $\sim 2\times 10^{-3}$ eV, respectively.
As $m_3$ increases from $\sim 2.5\times 10^{-3}$ eV to $\sim 5.0\times
10^{-2}$ eV, $\widetilde{m}_{2}$ continues to increase monotonically,
while $\widetilde{m}_{\tau}$ remains practically constant.}
$10^{-10}~{\rm eV} \leq m_3 \leq 5.0\times 10^{-2}$ eV.  This explains
why one can have successful leptogenesis for ${\rm min}(M_1) \ltap
5\times 10^{11}$ GeV even when $m_3 \cong 5\times 10^{-6}$ eV.  If
$R_{11} = 0$, we get for $m_3 \ll m_2$ and $R_{12} (R_{13})$ which
maximises the asymmetry $|\epsilon_{\tau}|$, as it follows from
eqs.~(\ref{IOtmtau1}) and (\ref{IOtm21}), $\widetilde{m}_{\tau} \sim
\widetilde{m}_{2} \sim m_3$.  Consequently, for $m_3 \ll 10^{-3}$ eV,
one also has $\widetilde{m}_{\tau},\widetilde{m}_{2} \ll 10^{-3}$ eV,
and for $M_1 < 10^{12}$ GeV the baryon asymmetry generated under these
conditions is strongly suppressed,
$|Y_B| \ll 8.6\times 10^{-11}$. \\

{\bf {\it B. Dirac CP-Violation in 
$U_{\rm PMNS}$ and Leptogenesis}}\\

We will assume that $\alpha_{21} = \alpha_{31} = 0$ (or $\alpha_{21} =
2\pi k$, $\alpha_{31} = 2\pi k'$, $k,k'=0,1,2,...$) and\footnote{If
  in the case of real $R_{1j}R_{1k}$ and, e.g.  $R_{11} = 0~(R_{12} =
  0)$, the Majorana phase $\alpha_{32(31)}$ entering into the
  expression for $\epsilon_\tau$ takes the CP-conserving value
  $\alpha_{32(31)} = \pi$, the CP-symmetry will be violated not only
  by the Dirac phase $\delta \neq k\pi$, $k=0,1,2,...$, but also by
  the matrix $R$ \cite{PPRio106}.  } 
analyse the simple possibility of
$R_{11} = 0$.  For $R_{11} = 0$ and $\alpha_{32} = 0$ we have:
$|\epsilon_\tau| \propto |{\rm Im}(U^{*}_{\tau 2}U_{\tau 3})| =
c^2_{23}c_{13}s_{12}s_{13}|\sin\delta| \ltap 0.054|\sin\delta|$, where
we have used $c^2_{23} =0.5$ and\footnote{Given the 2$\sigma$ allowed
  range of values of $c^2_{23} = (0.36 - 0.64)$, it is clear that the
  asymmetry $|\epsilon_\tau|$ can be bigger (smaller) by a factor
  $\sim 1.3~(0.72)$.}  $s_{12}\cong 0.55$.  Thus, for given $M_1$ the
maximal asymmetry $|Y_B|$ we can obtain will be smaller by a factor
$\sim (7 - 8)$ than the maximal possible asymmetry $|Y_B|$ in the
corresponding case of CP-violation due to the Majorana phase(s) in
$U_{\rm PMNS}$.  The wash-out mass parameter $\widetilde{m}_\tau$,
corresponding to $R_{12}$ maximising $|\epsilon_\tau|$, is given by
\begin{equation} 
\label{IOtmtauD1}
\widetilde{m}_\tau \cong  
\frac{m_2\,m_3}{m_3 + m_2}\,
\left [\left ( c_{12}s_{23} - \rho_{23}c_{13}c_{23} \right )^2 
+ s^2_{12}s^2_{13}c^2_{23}  
+ 2 s_{12}s_{13}c_{23}\, 
\left ( c_{12}s_{23} - \rho_{23}c_{13}c_{23} \right )\, \cos\delta 
\right ]\,,
\end{equation}
%
while $\widetilde{m}_2$ is determined by eq.~(\ref{IOtm21}).
Depending on the value of $\rho_{23}$, there are two quite different
cases to be considered.

If $\rho_{23} = -1$, the terms $s^2_{12}s^2_{13}c^2_{23}$ and $\propto
2s_{12}s_{13} c_{23}\cos\delta$ in the expression for
$\widetilde{m}_\tau$, eq.~(\ref{IOtmtauD1}), are subdominant and can
be neglected\footnote{The term $\propto 2s_{12}s_{13}
  c_{23}\cos\delta$, for instance, gives a relative contribution to
  $\widetilde{m}_\tau$ not exceeding 10\%.}.  
Thus,
$\widetilde{m}_\tau$ and $\widetilde{m}_2$ practically do not depend
on $\delta$ and we have for $c_{23} = s_{23} = 1/\sqrt{2}$:
$\widetilde{m}_\tau \cong 0.5(c_{12} + c_{13})^2 m_2m_3/(m_2 + m_3)
\cong 1.66m_2m_3/(m_2 + m_3)$, $\widetilde{m}_2\cong 0.34m_2m_3/(m_3 +
m_2)$.  Both the asymmetry $|\epsilon_\tau|$ and $|Y_B|$ are maximal
for $\delta = \pi/2 + k\pi$, $k=0,1,...$ The dependence of $|Y_B|$ on
$m_3$ is analogous to that in the case of CP-violation due to the
Majorana phase(s) in $U_{\rm PMNS}$: we have two similar maxima
corresponding to the CP-asymmetry being predominantly respectively in
the $\tau-$flavour
\begin{figure}[t!!]
\begin{center}
\vspace{-1.0cm}
\includegraphics[width=15cm,height=9cm]{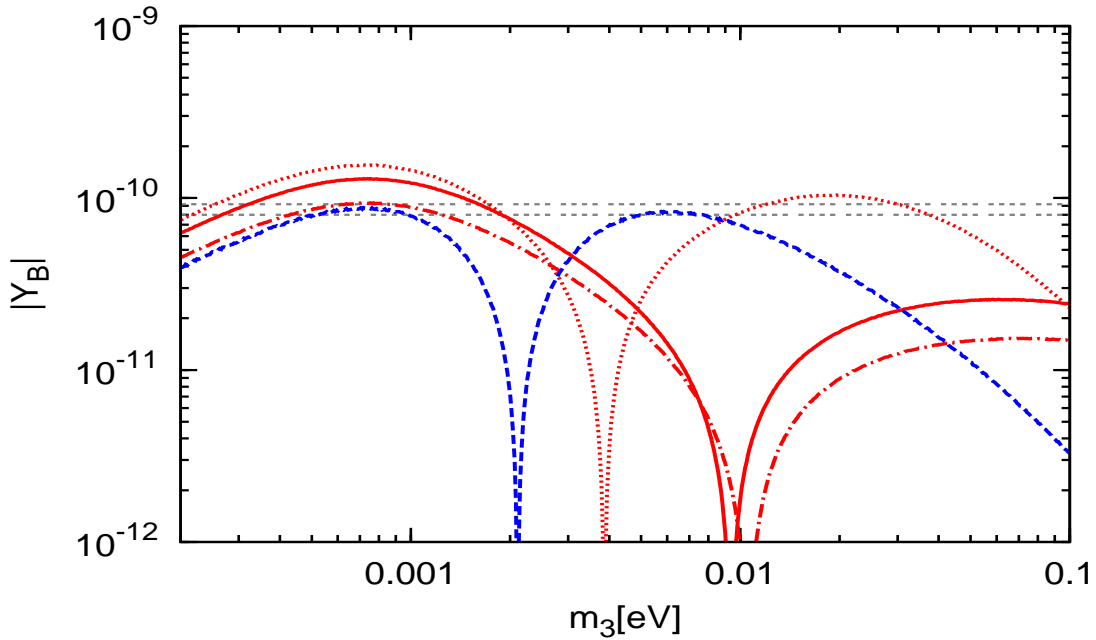}
\caption{\label{nTSDiracCP.eps} The dependence of $|Y_B|$ on $m_3$ in
  the case of spectrum with inverted ordering (hierarchy), real
  $R_{1j}R_{1k}$ and Dirac CP-violation, for $R_{11}=0$, $\delta=
  \pi/2$, $s_{13}=0.2$, $\alpha_{32} = 0$, $M_1=2.5\times 10^{11}~{\rm
    GeV}$ and ${\rm sgn}(R_{12}R_{13})= + 1~(-1)$ (red lines (blue
  dashed line)).  The baryon asymmetry $|Y_B|$ was calculated for a
  given $m_3$, using the value of $|R_{12}|$, for which the
  CP-asymmetry $|\epsilon_\tau |$ is maximal.  The results shown for
  ${\rm sgn}(R_{12}R_{13})= + 1$ are obtained for $\sin^2\theta_{23} =
  0.50;~0.36;~0.64$ (red solid, dotted and dash-dotted lines), while those
  for ${\rm sgn}(R_{12}R_{13})= -1$ correspond to $\sin^2\theta_{23} =
  0.5$. See text for further details.  }
\end{center}
\end{figure}
%
and in the $(e +\mu)-$flavour.  The two maxima are separated by a deep
minimum of $|Y_B|$ (Fig.~\ref{nTSDiracCP.eps}).  The maxima occur at
$m_3 \cong 7.5\times 10^{-4}$ eV ($|R_{12}| \cong 0.12$) and at $m_3
\cong 4.9\times 10^{-3}$ eV ($|R_{12}| \cong 0.30$), i.e. at values of
$m_3$ which differ by a factor\footnote{The positions of the two
  maxima and of the deep minimum of $|Y_B|$ under discussion exhibit
  very weak dependence on $\sin^2\theta_{23}$ when the latter is
  varied within its 2$\sigma$ allowed range, while the value of
  $|Y_B|$ at each of the two maxima changes according to $|Y_B|
  \propto \cos^2\theta_{23}$.}  
of $\sim 7$. At the first (second)
maximum we have $\eta(0.66\widetilde{m}_{\tau}) -
\eta(0.71\widetilde{m}_{2})\cong 0.044~(-0.046)$ and
\begin{equation}
|Y_B | \cong 3.5~(3.7)\times 10^{-13}\, |\,\sin \delta\,|\, 
\left(\frac{\sin\theta_{13}}{0.2}\right)\,
\left (\frac{\sqrt{|\deltaatm|}}{0.05~{\rm eV}}\right )
\left (\frac{M_1}{10^{9}~{\rm GeV}}\right )~.
\label{IHmaxYBD1}
\end{equation}
%
Thus, the measured value of $|Y_B|$, $8.0\times 10^{-11}\ltap
|Y_B|\ltap 9.2\times 10^{-11}$, can be reproduced for $M_1 \gtap
2.3~(2.2)\times 10^{11}~{\rm GeV}$.  The flavour effects in
leptogenesis are fully developed for $M_1 \ltap 5\times 10^{11}~{\rm
  GeV}$.  This upper bound and the requirement of successful
leptogenesis in the case of breaking of CP-symmetry due to the Dirac
phase in $U_{\rm PMNS}$, lead to the following lower limit on
$|\sin\theta_{13}\sin \delta|$ and thus on $\sin\theta_{13}$:
\begin{equation}
 |\sin\theta_{13}\sin \delta|\gtap 0.087\,,~~~ \sin\theta_{13}\gtap 0.087\,.
\label{IHminth13}
\end{equation}
%
The preceding lower bound corresponds to
\begin{equation}
|J_{\rm CP}|\gtap 0.02\,,
\label{IHminth13JCP}
\end{equation}
%
where $J_{\rm CP}$ is the rephasing invariant associated with the
Dirac phase $\delta$, which controls the magnitude of CP-violation
effects in neutrino oscillations\footnote{As is well-known, the
  Majorana phases, in contrast to the Dirac phase, do not affect the
  flavour neutrino oscillations \cite{BHP80,Lang87}.}
\cite{PKSP3nu88}.  Values of $s_{13}$ in the range given in 
eq.~(\ref{IHminth13}) can be probed in the forthcoming Double CHOOZ
\cite{DCHOOZ} and future reactor neutrino experiments\footnote{If we
  use $M_1 \ltap 10^{12}~{\rm GeV}$~\cite{davidsonetal,niretal,davidsonetal2} 
  for the maximal value of $M_1$ in this discussion, we get,
  obviously, $|\sin\theta_{13}\sin \delta|\gtap 0.044$,
  $\sin\theta_{13}\gtap 0.044$ and $|J_{\rm CP}|\gtap 0.01$.}
\cite{DayaB}.  CP-violation effects with magnitude determined by
$|J_{\rm CP}|$ satisfying (\ref{IHminth13JCP}) are within the sensitivity
of the next generation of neutrino oscillation experiments, designed
to search for CP- or T- symmetry violations in the oscillations
\cite{machines}. Since in the case under discussion the wash-out
factor $|\eta_B| \equiv |\eta(0.66 \widetilde{m}_\tau) -
\eta(0.71\widetilde{m}_2)|$ in the expression for $|Y_B|$ practically
does not depend on $s_{13}$ and $\delta$, while both $|Y_B| \propto
|s_{13}\sin\delta|$ and $|J_{\rm CP}| \propto |s_{13}\sin\delta|$,
there is a direct relation between $|Y_B|$ and $|J_{\rm CP}|$ for
given $m_3$ (or $m_2$) and $M_1$:
\begin{equation}
|Y_B | \cong 1.8\times 10^{-10}\,\, |J_{\rm CP}|\,|\eta_B|\,
\frac{m_2 - m_3}{\sqrt{|\deltaatm|}}\,
\left (\frac{\sqrt{|\deltaatm|}}{0.05~{\rm eV}}\right )
\left (\frac{M_1}{10^{9}~{\rm GeV}}\right )\,,
\label{IHmaxYBJCP1}
\end{equation}
%
where $\eta_B = \eta_B(m_2m_3/(m_2 + m_3),\theta_{12},\theta_{23})$
and we have used the best fit values of $\sin^2\theta_{12}$ and
$\sin^2\theta_{23}$.  In the case of IH spectrum we have $(m_2 -
m_3)/\sqrt{|\deltaatm|} \cong 1$ and $m_2m_3/(m_2 + m_3) \cong m_3$.
Similar relation between $|Y_B|$ and $|J_{\rm CP}|$ holds in an
analogous case of normal hierarchical light neutrino mass spectrum
\cite{PPRio106}.

We get somewhat different results if $\rho_{23} = +1$. Now there is a
strong compensation between the terms in the round brackets in the
expression (\ref{IOtmtauD1}) for $\widetilde{m}_\tau$ and we have
$\widetilde{m}_\tau \ll m_2m_3/(m_2 + m_3)$.  Correspondingly, one has
$\widetilde{m}_2 \cong 2m_2m_3/(m_2 + m_3) \gg \widetilde{m}_\tau$.
Thus, $\widetilde{m}_2$ practically does not depend on $\delta$ and on
the neutrino mixing angles.  The two wash-out mass parameters
$\widetilde{m}_2$ and $\widetilde{m}_\tau$ can differ by a factor
$\sim 100$. Indeed, for $s^2_{23} = c^2_{23} = 0.5$ and $s_{13}= 0.2$
and $s^2_{12}= 0.30$ one finds $\widetilde{m}_\tau/\widetilde{m}_2
\cong 0.5(0.0162 - 0.0156\cos\delta)$.  For fixed $\sin^2\theta_{12} =
0.30$, the magnitude of the ratio $\widetilde{m}_\tau/\widetilde{m}_2$
(which is practically independent of $m_2m_3/(m_2 + m_3)$) is very
sensitive to the value of $\theta_{23}$: for $s^2_{23}= 0.64$ we get
$\widetilde{m}_\tau/\widetilde{m}_2 \cong 0.5(0.0066 +
0.0043s^2_{13}/0.04 - 0.0107 (s_{13}/0.2) \cos\delta)$, while if
$s^2_{23}= 0.36$ one obtains $\widetilde{m}_\tau/\widetilde{m}_2 \cong
0.5(0.0794 + 0.0077s^2_{13}/0.04 - 0.0494 (s_{13}/0.2) \cos\delta)$.
The maxima of the asymmetry $|Y_B|$ take place at $\delta = \pi/2
+k\pi$, $k=0,1,2,...$.  For $\delta = \pi/2$, $s_{13}= 0.2$ and
$s^2_{23}= 0.64;~0.5;~0.36$ we get:
$\widetilde{m}_\tau/\widetilde{m}_2 \cong 0.52\times
10^{-2};~0.81\times 10^{-2};~4.36\times 10^{-2}$.  Therefore the two
maxima of $|Y_B|$ as a function of $m_3$, corresponding to the
CP-asymmetry being predominantly in the $(e +\mu)-$flavour and in the
$\tau-$flavour, can be expected to occur at values of $m_3$, which for
$s^2_{23}= 0.36;~0.5;~0.64$ and $s^2_{12} = 0.30$ would differ by a
factor of $\widetilde{m}_2/\widetilde{m}_\tau \sim 20;~120;~190$. The
position of the deep minimum of $|Y_B|$ between the two maxima would
also be very different for $s^2_{23}= 0.36$ and $s^2_{23}=
0.5~(0.64)$.  Obviously, the relative position on the $m_3$ axis of
two maxima and the minimum of $|Y_B|$ under discussion will depend not
only on the precise value of $\sin^2\theta_{23}$, but also on the
precise value of $\sin^2\theta_{12}$.

To be more concrete, the maximum of $|Y_B|$ (as a function of $m_3$),
associated with the CP-asymmetry being predominantly in the $(e
+\mu)-$flavour, takes place at $m_3 \cong 7.5\times 10^{-4}$ eV, i.e.
in the region of IH spectrum.  At this value of $m_3$,
$\eta(0.71\widetilde{m}_{2})$ is maximal, $\eta(0.71\widetilde{m}_{2})
\cong 0.068$, while $\eta(0.66\widetilde{m}_{\tau}) \ltap 0.005\ll
\eta(0.71\widetilde{m}_{2})$, and we have:
\begin{equation}
|Y_B | \cong 5.1\times 10^{-13}\, |\,\sin \delta\,|\, 
\frac{\sin\theta_{13}}{0.2}\,
\left (\frac{\sqrt{|\deltaatm|}}{0.05~{\rm eV}}\right )
\left (\frac{M_1}{10^{9}~{\rm GeV}}\right )~.
\label{IHmaxYBD2}
\end{equation}
%
The position of this maximum does not depend on $\theta_{12}$,
$\theta_{23}$, $\theta_{13}$ and $\delta$.  Thus, the measured value
of $|Y_B|$, $8.0\times 10^{-11}\ltap |Y_B|\ltap 9.2\times 10^{-11}$,
can be reproduced for a somewhat smaller value of $M_1 \gtap 1.6\times
10^{11}~{\rm GeV}$ than the corresponding value of $M_1$ we have found
for $\rho_{23} = -1$ (compare eqs.~(\ref{IHmaxYBD1}) and
(\ref{IHmaxYBD2})).  In the vicinity of the maximum there exists a
correlation between the values of $|Y_B|$ and $|J_{\rm CP}|$ similar
to the one given in eq.~(\ref{IHmaxYBJCP1}).  Now the requirement of
successful leptogenesis leads for $M_1 \ltap 5\times 10^{11}~{\rm
  GeV}$ to a somewhat less stringent lower limit on
$|\sin\theta_{13}\sin \delta|$, and thus on $\sin\theta_{13}$ and
$|J_{\rm CP}|$:
\begin{equation}
|\sin\theta_{13}\sin \delta|,~\sin\theta_{13}\gtap 0.063\,,~~~ 
|J_{\rm CP}| \gtap 0.015\,.
\end{equation}
%

The second maximum of $|Y_B|$, related to the possibility of the
CP-asymmetry being predominantly in the $\tau-$flavour, takes place,
as it is not difficult to convince oneself, at $m_2m_3/(m_2 +
m_3)\gtap 10^{-2}$ eV, i.e.  for values of $m_3 \gtap 1.2\times
10^{-2}$ eV in the region of neutrino mass spectrum with partial
inverted hierarchy.  In this case the factor in $|Y_B|$, which
determines the position of the maximum as a function of $m_3$, is
$((m_2 - m_3)/\sqrt{|\deltaatm|})\eta(0.66\widetilde{m}_{\tau})$,
rather than just $\eta(0.66\widetilde{m}_{\tau})$.  Taking this
observation into account, it is not difficult to show that for $\delta
=\pi/2$ and $s_{13}=0.2$ maximising $|Y_B|$, $s^2_{12}=0.30$ and, e.g.
$s^2_{23}=0.36~(0.50)$, the maximum occurs at $m_3 \cong
1.8~(5.0)\times 10^{-2}$ eV.  If $M_1 = 10^{11}$ GeV and
$\sqrt{|\deltaatm|}=5.0\times 10^{-2}$ eV, the value of $|Y_B|$ at
this maximum reads: $|Y_B| \cong 4.4~(1.1)\times 10^{-11}$.  For
$s^2_{23}=0.64$ we get for the same values of the other parameters
${\rm max}(|Y_B|) \cong 0.6\times 10^{-11}$.  Obviously, if $m_3\gtap
10^{-2}$ eV, the observed value of $|Y_B|$ can be reproduced for $M_1
\ltap 5\times 10^{11}$ GeV only if $s^2_{23} < 0.50$. The position of
the deep minimum of $|Y_B|$ at $m_3\gtap 10^{-3}$ eV, as we have
already indicated, is also very sensitive to the value of $s^2_{23}$:
for $\delta =\pi/2$, $s_{13}= 0.2$ and $s^2_{12}=0.30$, it takes place
at $m_3\cong 2\times 10^{-3}$ eV if $s^2_{23}=0.36$, and at $m_3\cong
10^{-2}$ eV in the case of $s^2_{23}=0.50$.  These features of the
dependence of $|Y_B|$ on $m_3$ are illustrated in Fig.~\ref{nTSDiracCP.eps}.

One can perform a similar analysis in the case of real $R_{1j}R_{1k}$,
$R_{12} = 0$ and $R_{11},R_{13} \neq 0$.  In this case we have
$|\epsilon_\tau| \propto |{\rm Im}(U^{*}_{\tau 1}U_{\tau 3})| =
c^2_{23}c_{13}c_{12}s_{13}|\sin\delta| \ltap 0.082|\sin\delta|$, and
 \begin{equation} 
 \label{IOtmtauD2}
 \widetilde{m}_\tau \cong  
 \frac{m_1\,m_3}{m_3 + m_1}\,
 \left [\left ( s_{12}s_{23} + \rho_{13}c_{13}c_{23} \right )^2 
 + c^2_{12}c^2_{23}s^2_{13}  
- 2 s_{13}c_{12}c_{23}\, 
\left ( s_{12}s_{23} + \rho_{13}c_{23}c_{13} \right )\, \cos\delta 
\right ]\,,
\end{equation}
%
$\widetilde{m}_2 = 2m_1m_3/(m_3 + m_1) - \widetilde{m}_\tau$, where
$\rho_{13}\equiv {\rm sgn}(R_{11}R_{13}) = \pm 1$ and $m_1 \cong m_2 =
\sqrt{m^2_3 +|\deltaatm|}$.  For $\rho_{13} = +1$, the two maxima of
$|Y_B|$ (as a function of $m_3$) have the same magnitude. They occur
at $\delta \cong 3\pi/4$, $s_{13}=0.2$ and $m_3 \cong 7.5\times
10^{-4}~(3.5\times 10^{-3})$ eV.  The maximal baryon asymmetry, ${\rm
  max}(|Y_B|)$, exhibits rather strong dependence on $s^2_{23}$.  For
$s^2_{23} = 0.36~(0.50)$, $M_1 = 5\times 10^{11}$ GeV and
$\sqrt{|\deltaatm|}=5.0\times 10^{-2}$ eV, we get ${\rm max}(|Y_B|)
\cong 1.7~(0.9) \times 10^{-10}$.  If $s^2_{23} > 0.50$, however, it
is impossible to reproduce the observed value of $|Y_B|$ for $M_1
\ltap 5\times 10^{11}$ GeV.  The same negative result holds for any
$s^2_{23}$ from the interval [0.36 - 0.64] if $s_{13} \ltap 0.10$.

In the case of $\rho_{13} = -1$, we have $|Y_B| \propto c^2_{23}$ in
the region of the maximum of $|Y_B|$ at $m_3 \cong 7.5\times 10^{-4}$
eV, associated with the CP-asymmetry being predominantly in the
$(e+\mu)-$ flavour. The baryon asymmetry $|Y_B|$ is maximal for
$\delta = \pi/2$ which maximises the CP-asymmetry $|\epsilon_\tau|$.
For $s_{13} = 0.2$, $c^2_{23} = 0.5$, $M_1 = 5\times 10^{11}$ GeV and
$\sqrt{|\deltaatm|}=5.0\times 10^{-2}$ eV, we find ${\rm max}(|Y_B|)
\cong 4.5 \times 10^{-10}$:
\begin{equation}
|Y_B | \cong 9.0\times 10^{-13}\, |\,\sin \delta\,|\, 
\frac{\sin\theta_{13}}{0.2}\,
\left (\frac{\sqrt{|\deltaatm|}}{0.05~{\rm eV}}\right )
\left (\frac{M_1}{10^{9}~{\rm GeV}}\right )~.
\label{IHmaxYBD3}
\end{equation}
%
Thus, the observed value of the baryon asymmetry can be reproduced for
relatively small values of $|\sin\theta_{13}\sin\delta|$, and
correspondingly of $\sin\theta_{13}$ and $|J_{\rm CP}|$:
\begin{equation}
|\sin\theta_{13}\sin \delta|,~\sin\theta_{13}\gtap 0.036\,,~~~ 
|J_{\rm CP}| \gtap 0.0086\,.
\label{IHmin2th13}
\end{equation}
%
In contrast, the position (with respect to $m_3$) of the maximum of
$|Y_B|$, associated with the CP-asymmetry being predominantly in the
$\tau-$flavour, and its magnitude, exhibit rather strong dependence on
$s^2_{23}$. For $s^2_{23}=0.36;~0.50;~0.64$, the maximum of $|Y_B|$ is
located at $m_3 \cong 0.7;~1.5;~3.0~\times 10^{-2}$ eV.  For $M_1
\ltap 5\times 10^{11}$ GeV, the measured value of $|Y_B|$, $8.0\times
10^{-11}\ltap |Y_B|\ltap 9.2\times 10^{-11}$, can be reproduced
provided $|\sin\theta_{13}\sin\delta| \gtap 0.046;~0.053;~0.16$ if
$s^2_{23}=0.36;~0.50;~0.64$.

%
\subsection{Neutrino Mass Spectrum with Normal Ordering}
%
%

We can get different results for light neutrino mass spectrum with
normal ordering.  The case of negligible $m_1$ and real
(CP-conserving) elements $R_{1j}$ of $R$ was analysed in detail
in~\cite{PPRio106}\footnote{Results for quasi-degenerate (QD)
  spectrum, $m_1 \cong m_2 \cong m_3$, $m_j^2 \gg |\dma|$, which
  requires $m_j \gtap 0.10$ eV, were also obtained in
  \cite{PPRio106}.}.  It was found that if the only source of
CP-violation is the Dirac phase $\delta$ in the PMNS matrix, the
observed value of the baryon asymmetry $Y_B \cong 8.6\times 10^{-11}$
can be reproduced if \cite{PPRio106} $|\sin\theta_{13}\sin\delta|
\gtap 0.09$.  Given the upper limit $|\sin\theta_{13}\sin\delta| <
0.2$, this requires $M_1 \gtap 2\times 10^{11}$ GeV.  The quoted lower
limit on $|\sin\theta_{13}\sin\delta|$ implies that we should have
$\sin\theta_{13} \gtap 0.09$ and that $|J_{\rm CP}| \gtap 2\times
10^{-2}$. The indicated values of $\sin\theta_{13}$ and of $|J_{\rm
  CP}|$ are testable in the reactor neutrino and long baseline
neutrino oscillation experiments under preparation
\cite{DCHOOZ,DayaB,machines,Future}.  If, however, the Dirac phase
$\delta$ has a CP-conserving value, $\delta \cong k\pi$,
$k=0,1,2,...$, and the requisite CP violation is due exclusively to
the Majorana phases $\alpha_{21,31}$ in $U$, the observed $Y_B$ can be
obtained for \cite{PPRio106} $M_1 \gtap 4\times 10^{10}$ GeV.  For
$M_1 = 5\times 10^{11}$ GeV, for which the flavour effects are fully
developed, the measured value of $Y_B$ can be reproduced for a rather
small value of $|\sin \alpha_{32}/2| \cong 0.15$, where $\alpha_{32}
\equiv \alpha_{31} - \alpha_{21}$.

In searching for possible significant effects of non-negligible $m_1$
in leptogenesis we have considered values of $m_1$ as large as 0.05
eV, $m_1 \leq 0.05$ eV.  For $3\times 10^{-3}~{\rm eV} \ltap m_1 \ltap
0.10$ eV, the neutrino mass spectrum is not hierarchical; the spectrum
exhibits partial hierarchy (see, e.g. \cite{BPP1}), i.e.  we have $m_1
< m_2 < m_3$.

We are interested in the dependence of the baryon asymmetry on the
value of $m_1$ .  In the case of neutrino mass spectrum with normal
ordering we have:
\begin{equation}
m_2 = \sqrt{m^2_1 + \deltasol}\,,~~~ 
m_3 = \sqrt{m^2_1 + \deltaatm}\,.
\label{m2m3PNH}
\end{equation}
%
We will illustrate the characteristic features of the possible effects
of $m_1$ in leptogenesis by analysing two simple possibilities:
$|R_{11}| \ll 1$ and $|R_{12}| \ll 1$.  Results of a more general
analysis performed without making a priori assumptions about the real
parameters $R_{11}$ and $R_{12}$ will also be presented.

We first set $R_{11} = 0$. The asymmetry $\epsilon_{\tau}$ is given
by:
\begin{equation}
\epsilon_\tau \, \cong \; -\, \frac{3M_1\sqrt{\deltaatm}}{16\pi v^2}\,
\left (\frac{m_3}{m_2} \right )^{\frac{1}{2}}\,
\frac{\sqrt{\deltaatm}}{m_2 + m_3} \, \rho_{23}\, r\,
{\rm Im}\left(U^{*}_{\tau 2}U_{\tau 3}\right)\,,
\label{NHepst1}
\end{equation}
%
where 
\begin{eqnarray}
\label{m2}
r\;=\;\frac{\left|R_{12} R_{13}\right|}
{|R_{12}|^2+\frac{m_3}{m_2} |R_{13}|^2}\,,
\label{NHr1}
\end{eqnarray}
%
and ${\rm Im}(U^{*}_{\tau 2}U_{\tau 3})$ is given in 
eq.~(\ref{IHIm23}).  The ratio in (\ref{NHr1}) is similar to the ratio in
eq.~(\ref{IHr1}).  Note, however, that the masses $m_{2,3}$ present in
eqs.~(\ref{IHepst1}) and (\ref{IHr1}) are very different from the
masses $m_{2,3}$ in eqs.~(\ref{NHepst1}) and (\ref{NHr1}).  Using
again the fact that $R^2_{12} + R^2_{13} = 1$ and the results given in
eq.~(\ref{IHmaxrR12}), it is easy to find that $r$ has a maximum for
\begin{equation}
R^2_{12} = \frac{m_3}{m_2 + m_3}\,,~~
R^2_{13} = \frac{m_2}{m_2 + m_3}\,,~~~R^2_{13} < R^2_{12}\,,
\label{maxrR12}
\end{equation}
%
where $m_2$ and $m_3$ are defined now by eq.~(\ref{m2m3PNH}).  At the
maximum
\begin{equation}
{\rm max}(r) = \frac{1}{2}\,
\left (\frac{m_2}{m_3} \right )^{\frac{1}{2}}\,.
\label{maxr1}
\end{equation}
%
For the value of $R_{12}$ ($R_{13}$), which maximises the ratio $|r|$
and correspondingly the asymmetry $|\epsilon_{\tau}|$, the relevant
wash-out mass parameters $\widetilde{m}_\tau$ and $\widetilde{m}_2$
are given by eqs.~(\ref{IOtmtau1}) and (\ref{IOtm21}) with $m_2$ and
$m_3$ determined by eq.~(\ref{m2m3PNH}). Since in the case of interest
$m_2 \gtap \sqrt{\deltasol} \cong 0.9\times 10^{-2}$ eV, $m_3 \gtap
\sqrt{\deltaatm} \cong 5.0\times 10^{-2}$ eV, we have $m_2m_3/(m_2 +
m_3) \gtap 0.7\times 10^{-2}$ eV.  The lightest neutrino mass $m_1$
can have any effect on the generation of the baryon asymmetry $Y_B$
only if $m^2_1 \gg \deltasol$ and if $m_1$ is non-negligible with
respect to $\sqrt{\deltaatm}$. Correspondingly, for non-negligible
$m_1$ of interest we will have $m_2m_3/(m_2 + m_3) \gtap 10^{-2}$ eV
and the baryon asymmetry will be generated in the ''strong wash-out''
regime, unless there is a strong cancellation between the first two
and the third terms in the expression for $\widetilde{m}_\tau$, see
eq.~(\ref{IOtmtau1}).
Obviously, the possibility of such a cancellation depends critically
on the ${\rm sgn}(R_{12}R_{13}) = \rho_{23}$.  It should also be clear
from eq.~(\ref{m2m3PNH}) and the dependence on $m_{2,3}$ of ${\rm
  max}(|\epsilon_{\tau}|)$ and of the corresponding
$\widetilde{m}_\tau$ and $\widetilde{m}_2$ (see eqs.~(\ref{NHepst1}),
(\ref{maxr1}), (\ref{IOtmtau1}) and (\ref{IOtm21})) that with the
increasing of $m_1$ beyond $\sim 10^{-2}$ eV the predicted
baryon asymmetry decreases.\\

{\bf{\it A. Leptogenesis due to 
Majorana CP-Violation in $U_{\rm PMNS}$}}\\

Suppose first that the Dirac phase $\delta$ in the PMNS matrix has a
CP-conserving value, $\delta = \pi\,k$, $k=0,1,2,...$, and that the
only source of CP-violation are the Majorana phases $\alpha_{21,31}$
in the PMNS matrix $U$.  In the specific case of $R_{11} = 0$ we are
considering the relevant CP-violating parameter is the difference of
the two Majorana phases $\alpha_{32} \equiv \alpha_{31} -
\alpha_{21}$.  In this case $|\epsilon_{\tau}| \propto {\rm
  Im}(U^*_{\tau 2}U_{\tau 3}) \cong c^2_{23}c_{12} |\sin
\alpha_{32}/2| \cong 0.42\,|\sin \alpha_{32}/2|$, where we have
neglected the possible subleading corrections due to terms
proportional to $\sin\theta_{13}$.  Even for $\sin\theta_{13} = 0.2$
the latter have practically no influence on the results we are going
to obtain and for simplicity we set $\sin\theta_{13} = 0$ in the
discussion of Majorana CP-violation which follows. For the wash-out
mass parameter $\widetilde{m}_\tau$ we find:
\begin{equation}
\widetilde{m}_\tau \cong m_2\, \frac{m_3}{m_2 + m_3}\, 
\left[ c^2_{12} s^2_{23} + c^2_{23} - 2\, \rho_{23}\,c_{23}\, s_{23}\, c_{12}\, 
\cos \frac{\alpha_{32}}{2}\right]\,.
\label{tmtau2}
\end{equation}
%
\begin{figure}[t!!]
\begin{center}
\vspace{-1cm}
\includegraphics[width=15cm,height=9.5cm]{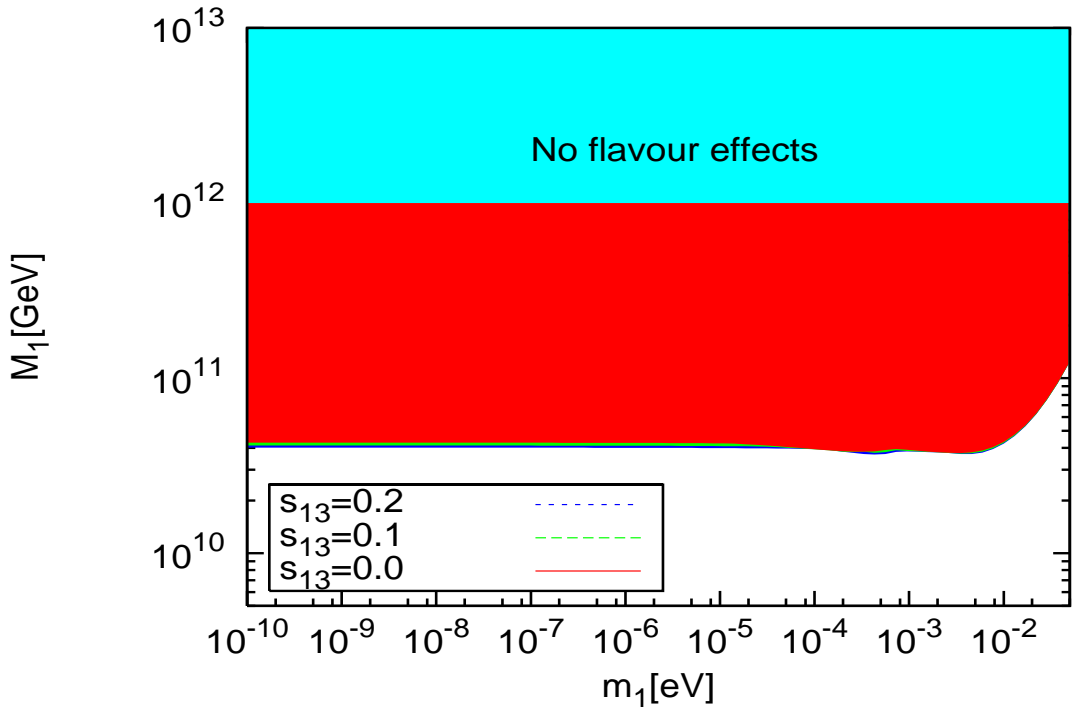}
\caption{
  \label{nfig1NOM1m1.eps} Values of $m_1$ and $M_1$ for which the
  ``flavoured'' leptogenesis is successful and baryon asymmetry $Y_B =
  8.6\times 10^{-11}$ can be generated (red shaded area). The figure
  corresponds to light neutrino mass spectrum with normal ordering.
  The CP-violation necessary for leptogenesis is due to the Majorana
  and Dirac phases in the PMNS matrix. The results shown are obtained
  using the best fit values of neutrino oscillation parameters:
  $\deltasol = 8.0\times 10^{-5}~{\rm eV}^2$, $\deltaatm = 2.5\times
  10^{-3}~{\rm eV}^2$, $\sin^2\theta_{12}=0.30$ and
  $\sin^22\theta_{23}=1$.  See text for further details.  }
\end{center}
\end{figure}
%
If $\cos \alpha_{32}/2 \cong 0$, the baryon asymmetry $Y_B$ is
produced in the strong wash-out regime and for $M_1 < 10^{12}$ GeV it
is impossible to reproduce the observed value of $Y_B \cong 8.6\times
10^{-11}$: the calculated asymmetry is too small. Actually, the
maximum of $|Y_B|$ in the case under discussion occurs, as can be
shown, for $\alpha_{32}\cong \pi/2 + \pi\, k$, $k=0,1,2,...$ There are
two distinctive possibilities to be considered corresponding to the
two possible signs of $\rho_{23}{\rm sgn}(\cos \alpha_{32}/2)$.  If
$\rho_{23}{\rm sgn}(\cos \alpha_{32}/2) = +1$, we have
$\widetilde{m}_\tau \cong 0.25 m_2m_3/(m_2 + m_3)$, the asymmetry in
the $\tau-$flavour ($(e +\mu)-$flavour) is produced in the weak
(strong) wash-out regime and we get for, e.g. $m_1 = 2\times
10^{-2}$~$(5\times 10^{-2}$) eV:
\begin{eqnarray}
  |Y_B|
  &\cong & 1.20\,(0.36)\times 10^{-12}
  \left (\frac{\sqrt{\deltaatm}}{0.05~{\rm eV}}\right )
  \left (\frac{M_1}{10^{9}~{\rm GeV}}\right )\,,~~~
  \alpha_{32}\cong \pi/2 + \pi\, k\,.
\label{maxYBMajCP}
\end{eqnarray}
%
Thus, for $m_1 = 2\times 10^{-2}$~ $(5\times 10^{-2}$) eV the measured
value of $Y_B \cong 8.6\times 10^{-11}$ can be obtained for $M_1 \gtap
7.2\times 10^{10}$~$(2.4\times 10^{11})$ GeV.

These results are illustrated in Fig.~\ref{nfig1NOM1m1.eps}, showing the
correlated values of $M_1$ and $m_1$ for which one can have successful
leptogenesis. The figure was obtained using the same general method of
analysis we have employed to produce Fig.~\ref{nfig1IOM1m3.eps}.  For given
$m_1$ from the interval $10^{-10} \leq m_1 \leq 0.05$ eV, a thorough
scan of the relevant parameter space was performed in the calculation
of $|Y_B|$, searching for possible non-standard features (enhancement
or suppression) of the baryon asymmetry.  The real elements $R_{1j}$
of interest of the matrix $R$, were allowed to vary in their full
ranges determined by the condition of orthogonality of $R$: $R^2_{11}
+ R^2_{12} + R^2_{13} = 1$. The Majorana and Dirac phases
$\alpha_{21,31}$ and $\delta$ were varied in the interval $[0,2\pi]$.
The calculations were performed again for three values of the CHOOZ
angle, $\sin\theta_{13} = 0;~0.1;~0.2$.  The relevant heavy Majorana
neutrino mass $M_1$ was varied in the interval $10^{9}~{\rm GeV} \ltap
M_1 \ltap 10^{12}$ GeV.  For given $m_1$, the minimal value of the
mass $M_1$, for which the leptogenesis is successful generating $Y_B
\cong 8.6\times 10^{-11}$, was obtained for the values of the other
parameters which maximise $|Y_B|$. The ${\rm min}(M_1)$ thus
calculated did not show any significant dependence on $s_{13}$.  For
$m_1 \ltap 7.5\times 10^{-3}$ eV we did not find any noticeable effect
of $m_1$ in leptogenesis: the results we have obtained practically
coincide with those corresponding to $m_1 = 0$ and derived in
\cite{PPRio106}.  The minimal value of $M_1 \cong 4\times 10^{10}$ GeV
seen in Fig.~1 corresponds to $R^2_{12} \cong 0.85$, $R^2_{13} \cong
0.15$ and $\alpha_{32}\cong \pi/2$ ($\rho_{23}{\rm sgn}(\cos
\alpha_{32}/2) = +1$); it does not depend on $m_1$.  For $7.5\times
10^{-3}~{\rm eV} \ltap m_1 \leq 5\times 10^{-2}$ eV, as our
calculations and Fig.~1 show, the predicted baryon asymmetry $Y_B$ for
given $M_1$ is generically smaller with respect to the asymmetry $Y_B$
one finds for $m_1 = 0$.  Thus, successful leptogenesis is possible
for larger values of ${\rm min}(M_1)$.  The corresponding suppression
factor increases with $m_1$ and for $m_1 \cong 5\times 10^{-2}$ eV
values of $M_1 \gtap 10^{11}$ GeV are required.

 If, however, $\rho_{23}{\rm sgn}(\cos \alpha_{32}/2) = -1$,
both the asymmetries in the $\tau-$flavour and in the 
$(e +\mu)-$flavour are generated under the conditions 
of strong wash-out effects. Correspondingly, 
it is impossible to have a successful leptogenesis 
for $M_1 < 10^{12}$ GeV if $m_1 \cong 5\times 10^{-2}$ eV.
If $m_1$ has a somewhat lower value, say 
$m_1 = 2\times 10^{-2}$ eV, the wash-out 
of the $(e +\mu)-$flavour asymmetry is less severe 
($\widetilde{m}_2 \cong 8.6\times 10^{-3}$ eV)
and the observed $Y_B$ can be reproduced 
for $\alpha_{32} = \pi/2 + \pi k$
if  $M_1 \gtap 2.5\times  10^{11}$ GeV. \\

{\bf{\it B. Leptogenesis due to Dirac CP-Violation 
in $U_{\rm PMNS}$}}\\

If the Majorana phases $\alpha_{21,31}$ have CP-conserving values and
the only source of CP-violation is the Dirac phase $\delta$ in $U_{\rm
  PMNS}$, one has $|\epsilon_{\tau}| \propto |c^2_{23}c_{13}s_{12}
s_{13}\sin \delta| \ltap 0.054|\sin \delta|$.  The factor
$c^2_{23}c_{13}s_{12} s_{13}$ in $|\epsilon_{\tau}|$ is smaller by at
least approximately an order of magnitude than the analogous factor
$c^2_{23}c_{12}$ in $|\epsilon_{\tau}|$ corresponding to Majorana
CP-violation we have considered above.  The study of this case of
Dirac CP-violation we have performed shows that as a consequence of
the indicated suppression it is impossible to generate the observed
value of the baryon asymmetry $|Y_B|$ for $M_1 \ltap 5\times 10^{11}$
GeV.

We have investigated also the possibility of $|R_{13}|$ being
sufficiently small, so that the term $\propto R_{11}R_{12}$ in the
expression for $\epsilon_{\tau}$ is the dominant one. We have found
that if $R_{13} = 0$, it is impossible to have successful leptogenesis
for $m_1 \ltap 0.05$ eV and $M_1 < 10^{12}$ GeV if the requisite
CP-violation is due to the Majorana and/or Dirac phases in $U_{\rm
  PMNS}$. The same conclusion is valid for normal hierarchical
spectrum, $m_1 \ll m_2 \cong \sqrt{\deltasol}$.\\

{\bf{\it C. The case of $R_{12} = 0$}}\\

\begin{figure}[t!!]
\begin{center}
\vspace{-1cm}
\includegraphics[width=15cm,height=8cm]{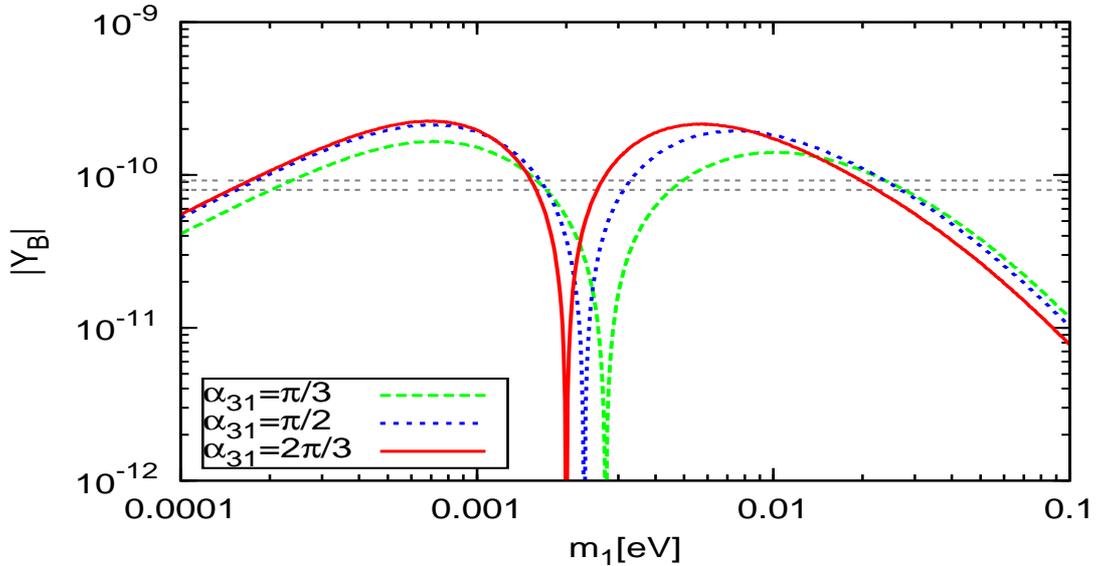}
\caption{The dependence of $|Y_B|$ on $m_1$ in the case of neutrino
  mass spectrum with normal ordering and real $R_{1j} R_{1k}$, for
  $R_{12}=0$, $s_{13}=0$, $M_1=1.5\times 10^{11}$ GeV and ${\rm
    sgn}(R_{11}R_{13}) = -1$. The red solid, the blue dotted, and the
  green dashed lines correspond to $\alpha_{31}=2\pi/3$, $\pi/2$, and
  $\pi/3$ respectively. The figure is obtained for
  $\theta_{23}=\pi/4$. \label{nN-R11.eps}}
\end{center}
\end{figure}

We get very different results if $R_{12}=0$, while $R_{11}R_{13} \neq
0$. In this case the expression for the asymmetry $\epsilon_{\tau}$
can be obtained formally from eq.~(\ref{NHepst1}) by replacing $m_2$
with $m_1$, $\rho_{23}$ with $\rho_{13}$, $U^{*}_{\tau 2}$ with
$U^{*}_{\tau 1}$ and the ratio $r$ with
\begin{equation}
r = \frac{\left| R_{11}\, R_{13} \right|}{|R_{11}|^2 + 
\frac{m_3}{m_1}\,|R_{13}|^2}\hspace{2mm},
\hspace{2mm}\hspace{2mm} R^2_{11} +  R^2_{13} = 1 \hspace{2mm}. 
\label{eq:sD2}
\end{equation}
%
As in the similar cases discussed earlier, the ratio $r$, and the
asymmetry $|\epsilon_{\tau}|$, take maximal values for
\begin{equation}
R^2_{11} = \frac{m_3}{m_1 + m_3} \hspace{2mm},\hspace{2mm}
R^2_{13} = \frac{m_1}{m_1 + m_3} \hspace{2mm},
\label{eq:maxrR11}
\end{equation}
%
and ${\rm max}(r) = 0.5 (m_1/m_3)^{\frac{1}{2}}$. The expression of
the asymmetry $|\epsilon_{\tau}|$ at the maximum (with respect to
$R_{11}$) reads:
\begin{equation}
  |\epsilon_\tau| \, \cong \;
  \frac{3M_1\sqrt{\deltaatm}}{32\pi v^2}\,
  \frac{\sqrt{\deltaatm}}{m_1 + m_3} \,  
  \left | {\rm Im}\left(U^{*}_{\tau 1}U_{\tau 3}\right)\right |\,.
\label{NHepst1R110}
\end{equation}
The wash-out parameters $\widetilde{m}_\tau$ and $\widetilde{m}_2$,
corresponding to the maximum $|\epsilon_\tau|$, are given by
\begin{eqnarray} 
\label{eq:tmtau2}
\widetilde{m}_\tau &\!\!=\!\!& \frac{m_1 \, m_3}{m_1 + m_3}\, 
\left[ \left| U_{\tau 1} \right|^2 +  \left| U_{\tau 3} 
\right|^2 + 2\, \rho_{13}\, {\rm Re}\left( 
U^*_{\tau 1}U_{\tau 3}\right) \right] \hspace{2mm}  \\[0.25cm]
&\!\!=\!\!& \frac{m_1 \, m_3}{m_1 + m_3}\,
\left (s^2_{12}\, s^2_{23} + c^2_{23} + 
2\, \rho_{13}\,c_{23}\, s_{23}\, s_{12}\, 
\cos \frac{\alpha_{31}}{2} \right ) \hspace{2mm}, 
\label{eq:tm2}
\end{eqnarray}
%
where we have set $s_{13} = 0$ in obtaining the second expression, and
$\widetilde{m}_2 = 2m_1m_3/(m_1 + m_3) - \widetilde{m}_\tau$.  Note
that if $m_1 \ll m_3 \cong 5\times 10^{-2}$ eV, the asymmetry
$|\epsilon_\tau|$ practically does not depend on $m_1$, while
$\widetilde{m}_{\tau,2} \sim {\cal O}(m_1)$.  This implies that the
dependence of ${\rm max}(|Y_B|)$ on $m_1$ as the latter increases,
will exhibit the same features as in the case of IH spectrum we have
already discussed: $|Y_B|$ will have two maxima, corresponding to the
CP-asymmetry being predominantly in the $\tau-$flavour and in the
$(e+\mu)-$flavour, separated with a deep minimum.  The analysis of the
similar case of IH spectrum suggests, that for $s_{13} = 0$, the
largest baryon asymmetry $|Y_B|$ is obtained for $\alpha_{31} \neq \pi
(2k + 1)$ and $\rho_{13} {\rm sgn}(\cos \alpha_{31}/2) = -1$.  These
features are confirmed by our numerical calculations and are
illustrated in Fig.~\ref{nN-R11.eps}.
%

In Fig.~\ref{nN-R11.eps} we have displayed the dependence of
$\left|Y_B\right|$ on $m_1$ in the case of neutrino mass spectrum with
normal ordering, $R_{12}=0$ and real $R_{11}R_{13}$.  The results
shown are obtained for $\rho_{13} = -1$, $\sin\theta_{13}=0$,
$M_1=3\times 10^{11}$ GeV, and three CP-violating values of the
Majorana phase $\alpha_{31}$, relevant for the calculation of $|Y_B|$:
$2\pi/3$; $\pi/2$; $\pi/3$.  The two maxima and the deep minimum of
$|Y_B|$ are evident in the figure.  The maximal values of $|Y_B|$ are
reached for $\alpha_{31} \cong 2\pi/3$.  Actually, in what concerns
the dependence of $|Y_B|$ on $\alpha_{31}$ and $\rho_{13}$ in the case
of $s_{13} = 0$, the following relation holds:
$|Y_B(\rho_{13},\alpha_{31})| = |Y_B(-\rho_{13},2\pi - \alpha_{31})|$.
The two maxima in question occur at $m_1 \cong 7.7\times 10^{-4}$~eV
and at $m_1\cong 5.5\times 10^{-3}$~eV.  At these maximum points we
have $\eta(0.66\widetilde{m}_{\tau})-\eta(0.71\widetilde{m}_2) \cong
(-0.044)~{\rm and}~ (+0.047)$, respectively.  The complete
compensation between $\eta(0.66\widetilde{m}_{\tau})$ and
$\eta(0.71\widetilde{m}_2)$, leading to $|Y_B| \cong 0$, takes place
at $m_1\sim 1.5\times 10^{-3}$ eV.  For $\alpha_{31} =2\pi/3$, the
baryon asymmetry at the two maxima reads:
\begin{equation}
 \left|Y_B\right|\cong 1.5~(1.4)\times 10^{-12}
\left(\frac{\sqrt{|\Delta m_A^2|}}{0.05~{\rm eV}}\right)
\left(\frac{M_1}{10^{9}~{\rm GeV}}\right)\;.
\end{equation}
%
Thus, one can have successful leptogenesis for $M_1 \gtap 5.3\times
10^{10}~{\rm GeV}$.

Similar analysis can be performed assuming that the only source of
CP-violation in leptogenesis is the Dirac phase $\delta$ in the
neutrino mixing matrix.  The results one obtains in this case are
analogous to those derived in the last part of Section 3.1 (see
eqs.~(\ref{IOtmtauD2}), (\ref{IHmaxYBD3}) and (\ref{IHmin2th13}) and
the related discussion).

%
\section{Conclusions}
%
%

In the present article we have investigated the dependence of the
``flavoured'' (thermal) leptogenesis on the lightest neutrino mass in
the case when the CP-violation necessary for the generation of the
observed baryon asymmetry of the Universe is due exclusively to the
Majorana and/or Dirac CP-violating phases in the PMNS neutrino mixing
matrix $U_{\rm PMNS}$. The two possible types of light neutrino mass
spectrum allowed by the data were considered: i) with normal ordering
($\deltaatm >0$), $m_1 < m_2 < m_3$, and ii) with inverted ordering
($\deltaatm < 0$), $m_3 < m_1 < m_2$.  The study was performed within
the simplest type I see-saw scenario with three heavy Majorana
neutrinos $N_j$, $j=1,2,3$, having a hierarchical mass spectrum with
masses $M_1 \ll M_{2,3}$.  Throughout this analysis we used the
``orthogonal'' parametrisation of the matrix of neutrino Yukawa
couplings, involving an orthogonal matrix $R$, $R^TR =
RR^T = {\bf 1}$.  The latter, in general, can be complex, i.e.
CP-violating.  In the present work we were primarily interested in the
possibility that $R$ is real, conserves the CP-symmetry, and the
violation of CP-symmetry necessary for leptogenesis is due exclusively
to the CP-violating phases in
$U_{\rm PMNS}$. In the case of hierarchical 
heavy Majorana neutrinos $N_{1,2,3}$, $M_1
\ll M_{2} \ll M_{3}$, the CP-violating lepton charge (flavour)
asymmetries $\epsilon_{l}$, $l=e,\mu,\tau$, relevant in
leptogenesis, are produced in the decays of the lightest one,
$N_1$. As a consequence, the generated baryon asymmetry $Y_B$ depends
(linearly) on the mass of $N_1$, $M_1$, and on the elements $R_{1j}$
of the matrix $R$, $j=1,2,3$, present in the neutrino Yukawa couplings
of $N_1$.

Our analysis was performed under the condition of negligible RG
running from $M_Z$ to $M_1$ of $m_j$ and of the parameters in $U_{\rm
  PMNS}$.  This condition is satisfied for sufficiently small values
of the lightest neutrino mass, ${\rm min}(m_j) \ltap 0.10$ eV. The
latter requirement is fulfilled for the normal hierarchical (NH), $m_1
\ll m_2 < m_3$, and inverted hierarchical (IH), $m_3 \ll m_1 < m_2$,
spectra, and for the spectra with partial hierarchy.  Therefore the
neutrino masses $m_j$, the solar, atmospheric and CHOOZ mixing angles,
$\theta_{12}$, $\theta_{23}$ and $\theta_{13}$, and the Majorana and
Dirac CP-violating phases, $\alpha_{21,31}$ and $\delta$, present in
$U_{\rm PMNS}$, are taken at the scale $\sim M_Z$, at which the
neutrino mixing parameters are measured.

We have investigated in detail the case of light neutrino mass
spectrum with inverted ordering and real (and CP-conserving) matrix
$R$, considering values of the lightest neutrino mass $m_3$ in the
range $10^{-10}~{\rm eV}\leq m_3 \leq 0.05~{\rm eV}$.  The case of IH
spectrum, $m_3 \ll m_1 < m_2$, $m_{1,2} \cong \sqrt{|\deltaatm|}$, is
of major interest since, for real elements $R_{1j}$ of $R$, IH
spectrum, $m_3 \cong 0$ and CP-violation due to the Majorana and/or
Dirac phases in $U_{\rm PMNS}$, it was shown \cite{PPRio106} to be
impossible to generate the observed baryon asymmetry $Y_B \cong
8.6\times 10^{-11}$ in the regime of ``flavoured'' leptogenesis (i.e.
for $M_1 \ltap 10^{12}~{\rm GeV}$): the resulting baryon asymmetry is
too small.  Therefore our investigation was focused primarily on the
effects in leptogenesis of a non-negligible $m_3$, having a value in
the range of the IH spectrum: $10^{-6}~{\rm eV} \ltap m_3 \ltap
5\times 10^{-3}~{\rm eV}$, $m_3 \ll m_{1,2} \cong \sqrt{|\deltaatm|}
\cong 5\times 10^{-2}~{\rm eV}$.  These effects can be particularly
large if $R_{11}\cong 0$ or $R_{12}\cong 0$.

We have found that in the case of IH spectrum with non-negligible
$m_3$, $m_3 \ll \sqrt{|\deltaatm|}$, the generated baryon asymmetry
$|Y_B|$ can be strongly enhanced in comparison with the asymmetry
$|Y_B|$ produced if $m_3 \cong 0$. The enhancement can be by a factor
of $\sim 100$, or even by a larger factor.  As a consequence, one can
have successful leptogenesis for IH spectrum with $m_3 \gtap 5\times
10^{-6}$ eV even if the elements $R_{1j}$ of $R$ are real and the
requisite CP-violation is provided by the Majorana or Dirac phase(s)
in the PMNS matrix.  The dependence of $|Y_B|$ on $m_3$ has the
following characteristic features.  As $m_3$ increases from the value
of $10^{-10}$ eV up to $10^{-4}$ eV, the maximal possible $|Y_B|$ for
a given $M_1$ increases monotonically, starting from a value which for
$M_1 \leq 10^{12}$ GeV is much smaller than the observed one, ${\rm
  max}(|Y_B|) \ll 8.6\times 10^{-11}$.  At $m_3 \cong {\rm few}\times
10^{-6}$ eV, we have ${\rm max}(|Y_B|) \cong 8.6\times 10^{-11}$ for
$M_1 \cong 5\times 10^{11}$ GeV.  As $m_3$ increases beyond ${\rm
  few}\times 10^{-6}$ eV, ${\rm max}(|Y_B|)$ for a given $M_1$
continues to increase until it reaches a maximum 
(Figs.~\ref{nfig1IOM1m3.eps}, ~\ref{nEMFig12YBm3.eps}
and~\ref{nTSDiracCP.eps}).  This maximum is located typically at $m_3
\cong (7.0 - 7.5)\times 10^{-4}$ eV.  It corresponds to the
CP-asymmetry being predominantly in the $(e +\mu)-$flavour.  As $m_3$
increases further, $|Y_B|$ rapidly decreases.  At certain value of
$m_3$, typically lying in the interval $m_3 \sim (1.5 - 2.5)\times
10^{-3}$ eV or $m_3 \sim (2.0 - 10.0)\times 10^{-3}$ eV, depending on
whether the CP-violation in leptogenesis is caused by the Majorana or
Dirac phases in $U_{\rm PMNS}$, $|Y_B|$ goes through a deep minimum:
one can have even $|Y_B| = 0$
(Figs.~\ref{nfig1IOM1m3.eps},~\ref{nEMFig12YBm3.eps}
and~\ref{nTSDiracCP.eps}).  This minimum of $|Y_B|$ corresponds to a
partial or complete cancellation between the asymmetries in the
$\tau-$flavour and in the $(e +\mu)-$flavour.  The position of the
minimum in the case of Dirac CP-violation from $U_{\rm PMNS}$ is very
sensitive to the precise value of the atmospheric and solar neutrino
mixing angles $\theta_{23}$ and $\theta_{12}$.  The dependence on
$\sin^2\theta_{23}$ is particularly strong
(Fig.~\ref{nTSDiracCP.eps}).  As $m_3$ increases further, $|Y_B|$
rapidly increases reaching a second maximum.  In magnitude the latter
is of the order of the first one in the case of Majorana CP-violation
from $U_{\rm PMNS}$. If the CP-violation is due to the Dirac phase in
$U_{\rm PMNS}$, the second maximum of $|Y_B|$ can be of the order of
the first one, or can be significantly smaller, depending on the
precise value of $\sin^2\theta_{23}$ (Fig.~\ref{nTSDiracCP.eps}).
This maximum corresponds to the CP-asymmetry being predominantly in
the $\tau-$flavour, rather than in the $(e +\mu)-$flavour.  For
Majorana (Dirac) CP-violation from $U_{\rm PMNS}$, the second maximum
of $|Y_B|$ occurs typically in the interval $m_3 \cong (4.0 -
8.0)\times 10^{-3}$ eV ($m_3 \cong (0.7- 7.0)\times 10^{-2}$ eV).  As
$m_3$ increases further, $|Y_B|$ decreases rather slowly
monotonically.  If CP-symmetry is violated by the Majorana phases in
$U_{\rm PMNS}$, we can have successful leptogenesis for $M_1\gtap
3.0\times 10^{10}~{\rm GeV}$.  A somewhat larger values of $M_1$ are
typically required if the CP-violation is due to the Dirac phase
$\delta$: $M_1\gtap 10^{11}~{\rm GeV}$. The requirement of successful
``flavoured'' leptogenesis in the latter case leads to the following
lower limits on $|\sin\theta_{13}\sin\delta|$, and thus on
$\sin\theta_{13}$ and on the rephasing invariant $J_{\rm CP}$ which
controls the magnitude of CP-violation effects in neutrino
oscillations: $|\sin\theta_{13}\sin \delta|,\sin\theta_{13}\gtap (0.04
- 0.09)$, $|J_{\rm CP}| \gtap (0.009 - 0.020)$, where the precise
value of the limit within the intervals given depends on the ${\rm
  sgn}(R_{11}R_{13})$ (or ${\rm sgn}(R_{12}R_{13})$) and on
$\sin^2\theta_{23}$.

The results we have obtained for light neutrino mass spectrum with
normal ordering, $m_1 < m_2 < m_3$, depend on whether $R_{11} \cong 0$
or $R_{12} \cong 0$.  If $R_{11} \cong 0$, we did not find any
significant enhancement of the baryon asymmetry $|Y_B|$, generated
within ``flavoured'' leptogenesis scenario with real matrix $R$ and
CP-violation provided by the neutrino mixing matrix $U_{\rm PMNS}$,
when the lightest neutrino mass was varied in the interval
$10^{-10}~{\rm eV}\leq m_1 \leq 0.05~{\rm eV}$: for $m_1 \ltap
7.5\times 10^{-3}~{\rm eV}$, the produced asymmetry $|Y_B|$
practically coincides with that corresponding to $m_1 = 0$
(Fig.~\ref{nfig1NOM1m1.eps}).  For $m_1 \gtap 10^{-2}~{\rm eV}$, the
lightest neutrino mass $m_1$ has a suppressing effect on the asymmetry
$|Y_B|$ (Fig.~\ref{nfig1NOM1m1.eps}).  If, however, $R_{12} \cong 0$,
the dependence of $|Y_B|$ on $m_1$ exhibits qualitatively the same
features as the dependence of $|Y_B|$ on $m_3$ in the case of neutrino
mass spectrum with inverted ordering (hierarchy) we have summarised
above: $|Y_B|$ possesses two maxima separated by a deep minimum
(Fig.~\ref{nN-R11.eps}).  Quantitatively, ${\rm max}(|Y_B|)$ is
somewhat smaller than in the corresponding IH spectrum cases.  As a
consequence, it is possible to reproduced the observed value of $Y_B$
if the CP-violation is due to the Majorana phase(s) in $U_{\rm PMNS}$
provided $M_1\gtap 5.3\times 10^{10}~{\rm GeV}$.

The results obtained in the present article show that the value of the
lightest neutrino mass in the cases of neutrino mass spectrum with
inverted and normal ordering (hierarchy) can have dramatic effect on
the magnitude of the baryon asymmetry of the Universe, generated
within the ``flavoured'' leptogenesis scenario with hierarchical heavy
Majorana neutrinos.

\section*{Acknowledgements}
 
This work was supported in part by the INFN under the program ``Fisica
Astroparticellare'', and by the Italian MIUR (Internazionalizzazione
Program) and Yukawa Institute of Theoretical Physics (YITP), Kyoto,
Japan, within the joint SISSA - YITP research project on ``Fundamental
Interactions and the Early Universe'' (S.T.P.).



\begin{thebibliography}{99}

\bibitem{FY} M.~Fukugita and T.~Yanagida,
Phys.\ Lett.\  B {\bf  174} (1986) 45.

\bibitem{kuzmin} V.A. Kuzmin, V.A. Rubakov 
and M.E. Shaposhnikov, 
Phys. Lett. B {\bf  155} (1985) 36.

\bibitem{LG1}
W.~Buchm\"uller, P.~Di Bari and M.~Pl\"umacher,
Nucl. Phys. B {\bf 643} (2002) 367;
Annals Phys.\  {\bf 315} (2005) 305.

\bibitem{LG2}
G.~F.~Giudice {\it et al.}, Nucl.\ Phys.\  B {\bf 685} (2004) 89.


\bibitem{others} H.~B.~Nielsen and Y.~Takanishi, 
Phys. Lett. B {\bf 507} (2001) 241; 
W.~Buchm\"uller and D.~Wyler, 
Phys.\ Lett.\ B {\bf 521} (2001) 291; 
J.~Ellis, M.~Raidal and T.~Yanagida, 
Phys. \ Lett.\ B {\bf 546} (2002) 228; 
S.~Davidson and A.~Ibarra, 
Nucl.\ Phys.\ B {\bf 648} (2003) 345.

\bibitem{others2} 
M.~Hirsch, S.~F.~King, 
Phys. Rev. D {\bf 64} (2001) 113005; 
G.C.\ Branco {\it et al.},  
Nucl.\ Phys.\ B {\bf 640} (2002) 202;
J. Ellis and M. Raidal,  Nucl.\ Phys.\ B {\bf 643} (2002) 229;
M.N. Rebelo, Phys. Rev. D {\bf 67} (2003) 013008.

\bibitem{PPR03} S. Pascoli, S.T. Petcov and W. Rodejohann, 
Phys. Rev. D {\bf 68} (2003) 093007.

\bibitem{PRST05}
S.T.~Petcov, W.~Rodejohann, T.~Shindou and Y.~Takanishi,
Nucl.\ Phys.\ B {\bf 739} (2006) 208.

\bibitem{PPRio106} S. Pascoli, S.T. Petcov and A. Riotto, 
Phys. Rev. D {\bf 68} (2003) 093007;
Nucl.\ Phys.\ B {\bf 739} (2006) 208.

\bibitem{BPont57} B.~Pontecorvo, 
                  Zh.\ Eksp.\ Teor.\ Fiz.\ {\bf 33} (1957) 549, 
{\bf 34} (1958) 247 and {\bf 53} (1967) 1717;
Z.~Maki, M.~Nakagawa and S.~Sakata, 
Prog.\ Theor.\ Phys.\  {\bf 28} (1962) 870.
%
\bibitem{BrancoJ06} 
 G.~C.~Branco, R.~Gonzalez Felipe and F.~R.~Joaquim,
  Phys.\ Lett.\  B {\bf 645} (2007) 432.

\bibitem{SBDibari06}
 S.~Blanchet and P.~Di Bari,
  JCAP {\bf 0703} (2007) 018.

\bibitem{Barbieri99}
R.~Barbieri, P.~Creminelli, A.~Strumia and N.~Tetradis,
Nucl. Phys. B {\bf 575} (2000) 61.

\bibitem{Nielsen02}
H.~B.~Nielsen and Y.~Takanishi,
Nucl.\ Phys.\  B {\bf 636} (2002) 305.

\bibitem{davidsonetal} 
A.~Abada {\it et al.},
JCAP {\bf 0604} (2006) 004.

\bibitem{niretal} 
E.~Nardi, Y.~Nir, E.~Roulet and J.~Racker,
JHEP {\bf 0601} (2006) 164.

\bibitem{davidsonetal2}
A.~Abada {\it et al.},
JHEP {\bf 0609} (2006) 010.

\bibitem{antusch} S.~Antusch, S.~F.~King and A.~Riotto,
JCAP {\bf 0611} (2006) 011.


\bibitem{seesaw}  P. Minkowski,
Phys.\ Lett.\  B {\bf  67} (1977) 421;
M. Gell-Mann, P. Ramond and R. Slansky, 
{\em Proceedings of the Supergravity Stony Brook Workshop}, 
New York 1979,  eds. P. Van Nieuwenhuizen and D. Freedman;
T. Yanagida,  
{\em Proceedinds of the Workshop on Unified Theories and Baryon Number in the
Universe},  Tsukuba, Japan 1979, ed.s A. Sawada and A. Sugamoto;
 R. N. Mohapatra and G. Senjanovic, Phys. Rev. Lett. {\bf 44} (1980) 912.

\bibitem{STPNu04} S.T.~Petcov,
{Nucl.\ Phys.\ B (Proc. Suppl.)}
{\bf 143} (2005) 159 (hep-ph/0412410).

\bibitem{MoscowH3Mainz}
V.~Lobashev {\it et al.},  
{Nucl.\ Phys.}  A {\bf 719}(2003) 153c;
%
K.~Eitel {\it et al.}, {Nucl. Phys. B (Proc. 
Suppl.)} {\bf 143} (2005) 197.
%

\bibitem{Hann06}  S.~Hannestad, H.~Tu and Y.~Y.~Y.~Wong,
  JCAP {\bf 0606} (2006) 025.

\bibitem{DiBGRaff06} 
  S.~Blanchet, P.~Di Bari and G.~G.~Raffelt,
  JCAP {\bf 0703} (2007) 012.

\bibitem{Casas01}
J.~A.~Casas and A.~Ibarra,
Nucl. Phys.  B {\bf  618} (2001) 171.

\bibitem{Casas07} J.~A.~Casas, A.~Ibarra and F. Jim\'enez-Albuquerque,
hep-ph/0612289.


\bibitem{rad} J.~A.~Casas, J.~R.~Espinosa, A.~Ibarra and I.~Navarro,
  Nucl.\ Phys.\  B {\bf 573} (2000) 652;
%
S.~Antusch {\it et al.}, Phys.\ Lett.\  B {\bf 519} (2001) 238;
%
T.~Miura, T.~Shindou and E.~Takasugi,
  Phys.\ Rev.\  D {\bf 66} (2002) 093002.

\bibitem{PST06} 
S.~T.~Petcov, T.~Shindou and Y.~Takanishi,
  Nucl.\ Phys.\  B {\bf 738} (2006) 219.

\bibitem{BPP1} 
 S.~M.~Bilenky, S.~Pascoli and S.~T.~Petcov,
  Phys.\ Rev.\  D {\bf 64} (2001) 113003.


\bibitem{BHP80} S.M. Bilenky, J. Hosek and S.T. Petcov,
             {Phys. Lett.} B {\bf 94} (1980) 495.

\bibitem{SchValle80Doi81} J. Schechter and J.W.F. Valle, 
{Phys. Rev.} D {\bf 22} (1980) 2227;
M.~Doi {\it et al.},
{Phys. Lett.} B {\bf 102} (1981) 323.

\bibitem{BCGPRKL2} A. Bandyopadhyay {\it et al.}, 
{ Phys.\ Lett.} B {\bf 608} (2005) 115, and 2005 (unpublished).

\bibitem{TSchwSNOW06} 
 T.~Schwetz,
  Phys.\ Scripta {\bf T127} (2006) 1.

\bibitem{Fogli06} G.L.~Fogli {\it et al.},
Prog. Part. Nucl. Phys. {\bf 57} (2006) 71.

\bibitem{CHOOZPV} M. Apollonio {\it et al.}, 
                 Phys. Lett. B {\bf 466} (1999) 415.

%
\bibitem{BiPet87} S.M. Bilenky and S.T. Petcov,
                {Rev. Mod. Phys.} {\bf 59} (1987) 67. 

\bibitem{STPFocusNu04} S.T.~Petcov, New J.\ Phys. {\bf 6} (2004) 109
({\it http://stacks.iop.org/1367-2630/6/109});
Physica Scripta {\bf T121} (2005) 94 (hep-ph/0504166); 
S.~Pascoli, S.T.~Petcov, hep-ph/0308034;
%
C.~Aalseth {\it et al.}, hep-ph/0412300;
A.~Morales and J.~Morales, 
{Nucl.\ Phys.\ B (Proc. Suppl.)} {\bf 114} (2003) 141.
%

\bibitem{Lang87} P. Langacker {\it et al.},  
{Nucl. Phys.} B {\bf 282} (1987) 589.

\bibitem{PKSP3nu88} P.I.~Krastev and S.T.~Petcov,
Phys.\ Lett.\ B {\bf 205} (1988) 84.


\bibitem{DCHOOZ}  
F.~Ardellier {\it et al.}  [Double Chooz Collaboration],
  hep-ex/0606025.

\bibitem{DayaB} See, e.g., K. M. Heeger, talk given at 
Neutrino'06 International Conference,
June 13 - 19, 2006, Sant Fe, U.S.A.

\bibitem{machines} See, for instance, 
 G.~De Lellis {\it et al.},
``Neutrino factories and superbeams,
Proceedings'', 7th International Workshop,
NuFact05, Frascati, Italy, June 21-26, 2005.


\bibitem{Future} 
C.~Albright {\it et al.}, physics/0411123;
Y.~Itow {\it et al.}, hep-ex/0106019;
D.~S.~Ayres {\it et al.}, hep-ex/0503053.
\end{thebibliography}
\end{document}